\documentclass[aps,prl,reprint,superscriptaddress]{revtex4-1}
\usepackage{amsmath}
\usepackage{amssymb}
\usepackage{graphicx}
\usepackage{xcolor} 
\usepackage{calc}

\usepackage{hyperref}
\hypersetup{pdfborder={0 0 0}} 

\begin{document}


\title{Adaptation controls synchrony and cluster states of coupled threshold-model neurons}



\author{Josef Ladenbauer}
\email[E-mail: ]{jl@ni.tu-berlin.de}
\affiliation{Institut f\"ur Softwaretechnik und Theoretische Informatik, Technische Universit\"at Berlin, Marchstra\ss{}e 23, 10587 Berlin, Germany}
\affiliation{Bernstein Center for Computational Neuroscience Berlin, Philippstra\ss{}e 13, 10115 Berlin, Germany}

\author{Judith Lehnert}
\thanks{These authors contributed equally to this work.}
\affiliation{Institut f{\"u}r Theoretische Physik, Technische Universit\"at Berlin, Hardenbergstra\ss{}e 36, 10623 Berlin, Germany}

\author{Hadi Rankoohi}
\thanks{These authors contributed equally to this work.}
\affiliation{Institut f\"ur Softwaretechnik und Theoretische Informatik, Technische Universit\"at Berlin, Marchstra\ss{}e 23, 10587 Berlin, Germany}
\affiliation{Bernstein Center for Computational Neuroscience Berlin, Philippstra\ss{}e 13, 10115 Berlin, Germany}

\author{Thomas Dahms}
\affiliation{Institut f{\"u}r Theoretische Physik, Technische Universit\"at Berlin, Hardenbergstra\ss{}e 36, 10623 Berlin, Germany}

\author{Eckehard Sch\"oll}
\affiliation{Institut f{\"u}r Theoretische Physik, Technische Universit\"at Berlin, Hardenbergstra\ss{}e 36, 10623 Berlin, Germany}

\author{Klaus Obermayer}
\affiliation{Institut f\"ur Softwaretechnik und Theoretische Informatik, Technische Universit\"at Berlin, Marchstra\ss{}e 23, 10587 Berlin, Germany}
\affiliation{Bernstein Center for Computational Neuroscience Berlin, Philippstra\ss{}e 13, 10115 Berlin, Germany}



\begin{abstract}  
We analyze zero-lag and cluster synchrony of delay-coupled non-smooth dynamical systems by extending the master stability approach, and apply this to networks of adaptive threshold-model neurons. 
For a homogeneous population of excitatory and inhibitory 
neurons we find (i) that subthreshold adaptation stabilizes or destabilizes synchrony 
depending on whether the 
recurrent synaptic excitatory or inhibitory couplings dominate, and (ii) that synchrony is always unstable for networks with balanced recurrent synaptic inputs. 
If couplings are not too strong, synchronization properties are similar for very different coupling topologies, i.e., random connections or spatial networks with localized connectivity.
We generalize our approach for two subpopulations of neurons with non-identical local dynamics, including bursting, for which 
activity-based adaptation controls the stability of cluster states, independent of a specific coupling topology.
\end{abstract}


\maketitle





\section{I. Introduction}
Synchronization and cluster states in complex networks have been in the focus
of intense research in physics, biology, neuroscience, and technology
\cite{Strogatz2001, Boccaletti2006, Arenas2008,
Engel2001, Schnitzler2005, Uhlhaas2010, Wang2010}. The key question is how the 
properties of the individual elements, the type of the coupling, and the topology
of connections determine the collective dynamics.

A powerful technique to address this question is the master stability function
(MSF) formalism \cite{Pecora1998}, which has been developed for smooth coupled
dynamical systems to analyze full synchrony 
\cite{Barahona2002, Nishikawa2003, Chavez2005, Dhamala2004} and cluster
states with synchronized groups of elements
\cite{SOR07, Choe2010, SEL12, WIL13, DAH12}. 
In various applications, however,
non-smooth dynamical systems arise \cite{Bernardo2008}. Examples include
electronic circuits \cite{Bernardo1998,Banerjee2001}, hybrid
control systems
\cite{Ye1998,Cassandras2001}, and biological networks
\cite{Battogtokh2006a, Aihara2010, Izhikevich2007}. Specifically in neuroscience,
theoretical investigations of neuronal network activity often involve neuron models
of the integrate-and-fire (IF) type, which include a discontinuity and
non-smooth models of synaptic couplings between them. IF models
are commonly used in computational studies of network activity
\cite{Izhikevich2008, Vogels2009, Litwin-Kumar2012, Destexhe2009, Brunel2000, Gigante2007c, Augustin2013}, 
and form the hardware-elements of neuromorphic systems
designed for spike-based, brain-style computations \cite{Indiveri2006,
Jo2010}. In this contribution we develop an MSF formalism for delay-coupled non-smooth
dynamical systems, and we demonstrate its potential by studying synchrony
and cluster states for recurrent networks of adaptive IF model neurons.

\section{II. Network model}
Consider a network of $N$ IF neurons which are coupled by delayed
synaptic currents. We choose the adaptive exponential
integrate-and-fire (aEIF) model \cite{Brette2005}, which is a two-variable neuron model that can well reproduce a variety of subthreshold dynamics and spike patterns
observed in cortical neurons \cite{[{The aEIF neuron model employed here uses dimensionless variables.
It is a rescaled variant of the original model, but with fewer parameters, see }] Touboul2008,
Naud2008, Jolivet2008}. The subthreshold dynamics of each aEIF neuron is
given by ($i=1,...,N$):
\begin{align}
&\dot{V}_i = -V_i + \mathrm{exp}(V_i) - w_i + I + \lambda \sum_{j=1}^{N} c_{ij} s_{j}(t-\tau), \label{eq_neuralnetwork_1}
\end{align}
\begin{align}
&\dot{w}_i = \frac{a V_i - w_i}{\tau_w}, \qquad \quad
\dot{s}_i = -\frac{s_i}{\tau_s}, \label{eq_neuralnetwork_3}
\end{align}
where $V_i$ is the membrane voltage, $w_i$ is the strength of the adaptation current, and $s_i$ is the strength of an effective synaptic (output) current. Equation~\eqref{eq_neuralnetwork_1} effectively describes the change of the membrane voltage
of each neuron in reaction to the ionic currents which flow through its membrane. Five currents are taken into account (Eq.~\eqref{eq_neuralnetwork_1}, left to right): The leak current, the fast sodium current at spike initiation \footnote{The fast sodium current of the aEIF model includes the effect of the sodium current, which is responsible for the generation of spikes.}, 
an adaptation current, which reflects slowly varying,
activity-dependent potassium currents, an external driving input $I$, and the weighted sum of time-delayed synaptic input currents caused by the other neurons in the network with overall strength $\lambda$, weights $c_{ij}$, and delay $\tau$. Equation~\eqref{eq_neuralnetwork_3}, left, quantifies, how the strength of the adaptation current depends on the membrane potential, where $a \geq 0$ and $\tau_w$ is the adaptation time constant. If a neuron is activated, the adaptation current increases, counteracting this activation. Equation~\eqref{eq_neuralnetwork_3}, right, describes the dynamics of the synaptic current. The activation of a model synapse decays exponentially with relaxation time $\tau_s$.

When the membrane potential increases beyond a threshold value $V_{th}$, an action potential (spike)
is generated. IF models do not describe the dynamics of the action potential
explicitly. Instead, the synaptic output current is incremented by 1 (indicating the spike) 
\colorbox{yellow!25}{
\parbox[!t]{\linewidth-10pt}{
\leftskip3pt\rightskip7pt
\begin{center} \small{\textbf{AUTHOR SUMMARY}} \end{center}
\hspace{7pt} \small{How synchronous states of complex networks depend on properties of the network nodes, their coupling, and the pattern of connections is of great interest across several sciences; applications range from coupled lasers to biological networks. A powerful method to address this question is the master stability function formalism, which can be used to efficiently predict the stability of zero-lag and cluster synchrony for a wide range of connection patterns. Here we extend this formalism to coupled non-smooth dynamical systems, which, for example, occur in electronic circuits, hybrid control systems, biology, and particularly in neuroscience, where non-smooth network models have a long history.}

\hspace{7pt} \small{Applying our formalism to cortical network models we identify activity-driven adaptation (a node property) as a key factor which determines the stability of synchrony largely independent of the specific pattern of connections. In particular, if couplings are not too strong, the level of adaptation that guarantees stability of synchrony in a model of a local cortical column (random connectivity) also predicts stability of global synchronous activity in a model of a cortical region (spatially structured connectivity) and vice versa. This is interesting because the adaptation properties of neurons can be changed by the neuromodulatory systems of the brain and thus provide a mechanism to control the dynamics of cortical networks.}
\vskip7pt}} \vskip10pt
\noindent and the
membrane potential is instantaneously reset to a lower value $V_r$. For the aEIF neuron, the
strength of the adaptation current is also changed after the spike has occurred. It is
increased by a value of $b \geq 0$
implementing the mechanism of spike-based adaptation. Therefore, the aEIF model takes two
mechanisms for activity-based adaptation into account, both of which are common to real
cortical neurons: A subthreshold mechanism, driven by the increase of the membrane potential,
and a spike-based mechanism, which is activated every time a spike has occurred. 
It has been shown that subthreshold adaptation can induce synchrony for pairs of symmetrically coupled excitatory neurons \cite{Ermentrout2001, Ladenbauer2012} and spike-dependent adaptation can cause networks to split up into phase locked clusters \cite{Kilpatrick2011}. Here we examine under what conditions the two adaptation mechanisms can (de)stabilize synchrony and different cluster states for a wide range of biologically plausible patterns of connections. 

The network of aEIF model neurons, which we have introduced above, is a typical example of 
a network of non-smooth dynamical systems with time delays. These networks can be described in general by
two sets of equations, one for the dynamics of the network elements between the discontinuities,
\begin{equation}
\dot{\mathbf{x}}_{i}^{\hphantom{+}} \!= \mathbf{f}(\mathbf{x}_i) + \sum_{j=1}^{N} c_{ij} \mathbf{h}(\mathbf{x}_{j,\tau}) \qquad \text{if} \; \varphi(\mathbf{x}_i) \neq 0,
\label{eq_network_comp1}
\end{equation}
and one for the jumps,
\begin{equation}
\mathbf{x}_i^+ \!= \mathbf{g}(\mathbf{x}_i) \qquad \text{if} \; \varphi(\mathbf{x}_i) = 0, \label{eq_network_comp2}
\end{equation}  
$i = 1,\dots,N$, $\mathbf{x}_i \in \mathbb{R}^m$ and  $\mathbf{x}_{i,\tau} \equiv \mathbf{x}_i(t-\tau)$. 
We assume that $\mathbf{x}_i(t)$ is piecewise continuous and that $\mathbf{f}$,  $\mathbf{g}$, $\mathbf{h}$ are differentiable vector functions \footnote{Note that $\mathbf{h}$ does not depend on $\mathbf{x}_i$ in Eq.~\eqref{eq_network_comp1}. The following derivation, however, can be extended to coupling functions $\mathbf{h}(\mathbf{x}_i,\mathbf{x}_{j,\tau})$ in a straightforward way}.
$\varphi$ is a scalar-valued, differentiable function that indicates the occurrence of a discontinuity, and 
\mbox{$\mathbf{x}_i^+(t) \equiv \mathrm{lim}_{s \searrow t} \, \mathbf{x}_i(s)$} denotes a right-sided limit. 

\section{III. Master Stability Function for full synchrony}
We now extend the MSF formalism to non-smooth dynamical systems as
described by Eqs.~\eqref{eq_network_comp1} and \eqref{eq_network_comp2}. We assume constant row sum of the coupling matrix, $\sum_{j=1}^{N} c_{ij} = \bar{c}$ $\forall i$.
Then a zero-lag synchronous state exists, in which every element of the network evolves according to the same
equation \mbox{$\dot{\mathbf{x}} = \mathbf{f}(\mathbf{x}) + \bar{c} \, \mathbf{h}(\mathbf{x}_{\tau}) \equiv \mathbf{F}(\mathbf{x},\mathbf{x}_{\tau})$}. We denote this solution by $\mathbf{x}_s$ and the 
set of times at which $\mathbf{x}_s$ changes discontinuously by $\lbrace t_s \rbrace$.

\begin{figure*}[!htp]
\includegraphics[width=0.75\linewidth]{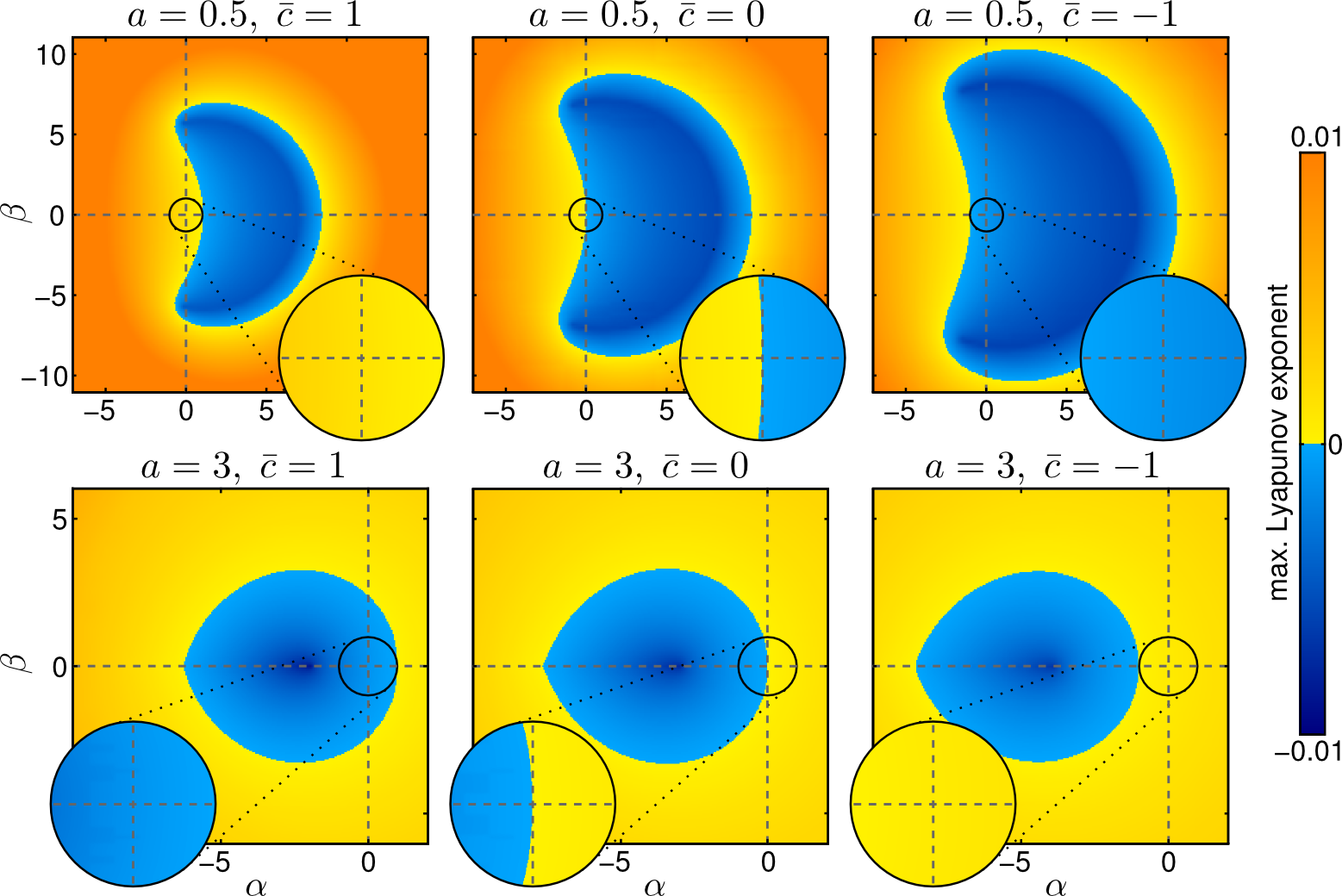} 
\caption{(Color online) Stability of synchrony for a homogenous population of coupled aEIF neurons.
The maximum Lyapunov exponent (color bar) is plotted as a function of the real ($\alpha$) and
imaginary ($\beta$) parts of the eigenvalues of the coupling matrix $\mathbf{C}$.
Different panels show the results for strong (top) and weak (bottom) subthreshold
adaptation, and for the excitation dominated (left), balanced (center), and inhibition
dominated (right) regimes. Parameters were selected such that the dynamics of the aEIF model closely matches the regular spiking dynamics of cortical neurons \cite{Naud2008, Destexhe2009}:
$\lambda = 5$, $\tau = 0.3$, $\tau_w = 10$, $\tau_s = 0.3$, $V_{th} = 5$, $V_r = -5$, $b = 2.5$.
The external input $I$ was chosen such that the oscillation
period was always equal to $5$. Circles indicate the unit disc in the eigenvalue plane (insets show blow-up). 
\label{fig1}}
\end{figure*}
We next assess the stability of the fully synchronous solution $\mathbf{x}_s$ by linearising 
Eqs.~\eqref{eq_network_comp1} and \eqref{eq_network_comp2} around $\mathbf{x}_s$. Between the discontinuities at $t \in \lbrace t_s \rbrace$ and the
corresponding kinks of $\mathbf{x}_s$ at $t \! - \! \tau \in \lbrace t_s \rbrace$, we obtain the variational equation
\begin{equation}
\dot{\underline{\boldsymbol{\xi}}} = \left[ \mathbf{I}_N \otimes D \mathbf{f}(\mathbf{x}_s) \right] \underline{\boldsymbol{\xi}} + \left[ \mathbf{C} \otimes
D \mathbf{h}(\mathbf{x}_{s,\tau}) \right] \underline{\boldsymbol{\xi}}_{\tau}
\label{eq_variational_xi1}
\end{equation}  
for the perturbations $\underline{\boldsymbol{\xi}} \equiv (\boldsymbol{\xi}_1,\dots,\boldsymbol{\xi}_N)^T \in \mathbb{R}^{Nm}$ from the synchronous solution (cf. \cite{Dhamala2004}), where $\underline{\boldsymbol{\xi}}_{\tau} \equiv \underline{\boldsymbol{\xi}}(t-\tau)$, $\mathbf{I}_N$ is the $N$-dimensional unity matrix, $\mathbf{C}$ is the coupling matrix, and $\otimes$ denotes the Kronecker product. For non-smooth dynamical systems, however, the variational equation must be complemented by the appropriate linearized transition conditions at the times $t, t-\tau \in \lbrace t_s \rbrace$. 
Using first order approximations of the times at which $\varphi(\mathbf{x}_s + \boldsymbol{\xi}_i) = 0$
and using Taylor expansions of
\mbox{$\mathbf{f}(\mathbf{x}_i) + \sum_{j=1}^{N} c_{ij} \mathbf{h}(\mathbf{x}_{j,\tau})$} around
$\mathbf{x}_s$ and $\mathbf{x}^+_s$, at both $t_s$ and $t_s + \tau$ we finally obtain (see Appendix I for a derivation):
\begin{align} 
\underline{\boldsymbol{\xi}}^+ &= \left[ \mathbf{I}_N \otimes \mathbf{A} \right] \underline{\boldsymbol{\xi}} & &t \in \lbrace t_s \rbrace , \label{eq_variational_xi2} \\
\underline{\boldsymbol{\xi}}^+ &= \underline{\boldsymbol{\xi}} + \left[ \mathbf{C} \otimes
\mathbf{B} \right] \underline{\boldsymbol{\xi}}_{\tau} & &t \! - \! \tau \in \lbrace t_s \rbrace .
\label{eq_variational_xi3}
\end{align}
The matrices $\mathbf{A}$ and $\mathbf{B}$ are given by:
\begin{align} 
\mathbf{A} &\equiv D \mathbf{g}(\mathbf{x}_s) \! + \! \frac{\left( \mathbf{F}(\mathbf{x}_s^+, \mathbf{x}_{s,\tau}) \! - \! D \mathbf{g}(\mathbf{x}_s) \mathbf{F}(\mathbf{x}_s, \mathbf{x}_{s,\tau}) \right) \! D \varphi(\mathbf{x}_s)}{D \varphi(\mathbf{x}_s) \mathbf{F}(\mathbf{x}_s, \mathbf{x}_{s,\tau})} \label{eq_transition_matrix1} \\
\mathbf{B} &\equiv \frac{\left( \mathbf{F}(\mathbf{x}_s, \mathbf{x}_{s,\tau}^{+}) - \mathbf{F}(\mathbf{x}_s, \mathbf{x}_{s,\tau}) \right) D \varphi(\mathbf{x}_{s,\tau})}{D \varphi(\mathbf{x}_{s,\tau}) \mathbf{F}(\mathbf{x}_{s,\tau}, \mathbf{x}_{s,2\tau})}. \label{eq_transition_matrix2}
\end{align}
Block-diagonalization of Eqs.~\eqref{eq_variational_xi1}--\eqref{eq_variational_xi3} then leads to the
master stability equations
\begin{align}
\dot{\boldsymbol{\zeta}}^{\hphantom{+}} &\!= D \mathbf{f}(\mathbf{x}_s) \boldsymbol{\zeta} \! + \! \left( \alpha \!+ \! i \beta \right) \! D \mathbf{h}(\mathbf{x}_{s,\tau}) \boldsymbol{\zeta}_{\tau} & &\!\!\! t, t \! - \! \tau \notin \lbrace t_s \rbrace , \label{eq_mse1} \\ 
\boldsymbol{\zeta}^+ &\!= \mathbf{A} \boldsymbol{\zeta} & &\!\! t \in \lbrace t_s \rbrace , \label{eq_mse2} \\
\boldsymbol{\zeta}^+ &\!= \boldsymbol{\zeta} + \left( \alpha \! + \! i \beta \right) \mathbf{B} \boldsymbol{\zeta}_{\tau} & &\!\! t \! - \! \tau \in \lbrace t_s \rbrace , \label{eq_mse3}
\end{align} 
where $\boldsymbol{\zeta} \equiv \left( \mathbf{z}^T \otimes \mathbf{I}_m \right) \underline{\boldsymbol{\xi}}$, and $\mathbf{z}$ is the normalized eigenvector of $\mathbf{C}$ that corresponds to the eigenvalue $\alpha + i \beta$. Hereby we  separate the variational equation for perturbations in the longitudinal direction ($\alpha + i \beta = \bar{c}$) from those in the transverse directions. We can now calculate the largest Lyapunov exponent for the solution $\boldsymbol{\zeta} \equiv \mathbf{0}$ as a function of $\alpha$ and $\beta$, which defines the MSF. Here we apply the numerical scheme proposed in \cite{Farmer1982} to obtain the Lyapunov exponents for Eqs.~\eqref{eq_mse1}--\eqref{eq_mse3}.

\section{IV. Stability of full synchrony for networks of adaptive neurons}
Let us first apply our MSF formalism to a homogeneous population of coupled aEIF neurons which are
driven by a constant external input $I$. 
Figure~\ref{fig1}
shows the result for six different sets of parameters. The three columns from left to right show the MSF for networks with dominant excitatory couplings (row sum $\bar{c} = 1$), balanced excitation and inhibition ($\bar{c} = 0$), and dominant inhibitory couplings ($\bar{c} = -1$), respectively. The two rows show the MSF for neurons whose subthreshold adaptation is weak
(top) and strong (bottom), respectively. 
If the eigenvalues of the coupling matrix lie within the unit circle (highlighted by insets in Fig.~\ref{fig1}), stable zero-lag synchrony is predicted for excitation dominated networks
of neurons with strong subthreshold adaptation and for inhibition
dominated networks of neurons with weak subthreshold adaptation. An increase of spike-triggered adaptation (larger $b$) on the other hand does not stabilize synchrony in excitation dominated networks (not shown). For balanced networks synchrony remains unstable independently
of the choice of adaptation parameters, because there are negative and positive values of the largest Lyapunov exponent in any small neighborhood of the origin 
\footnote{We observed this behavior for a wide range of model parameters}.
In summary, we identify neuronal
adaptation, which in real cortex is under top-down control of the brain's neuromodulatory systems 
\cite{McCormick1992, Abbott2000, Destexhe2004}, as
one of the key factors for stabilizing or destabilizing synchrony in recurrent networks.

\begin{figure*}[!htp]
\includegraphics[width=0.75\linewidth]{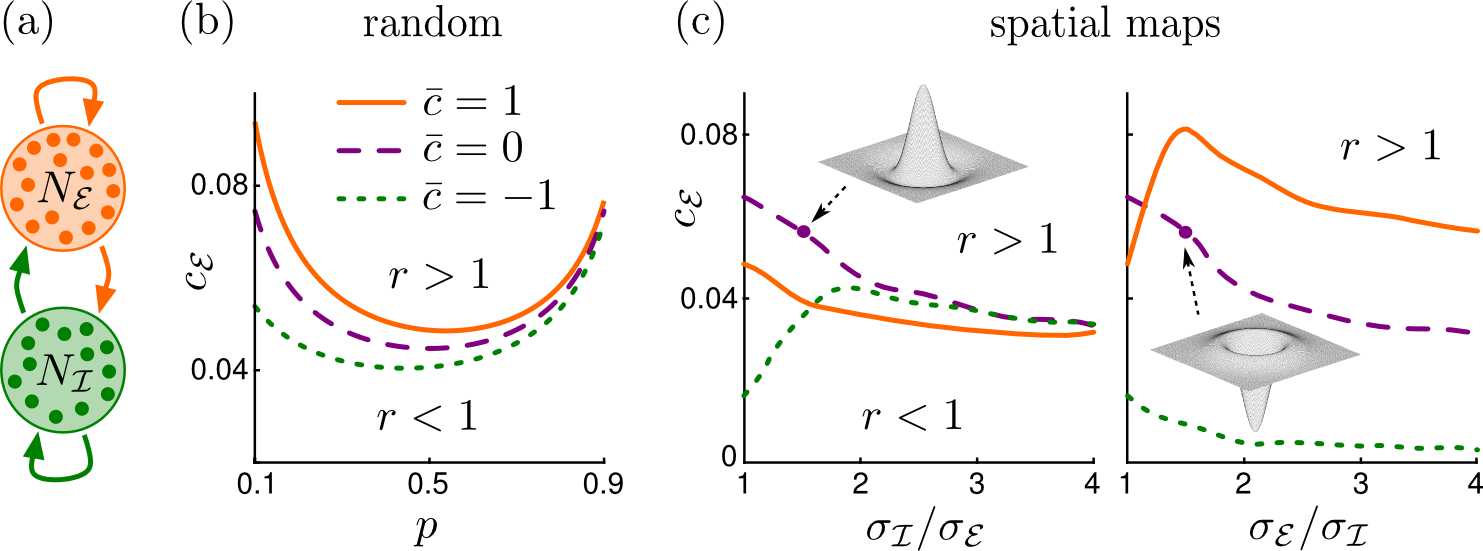} 
\caption{(Color online) Eigenvalue spectra of coupling matrices for two biologically relevant connectivity schemes. 
(a) Cartoon of the network: two subpopulations of $N_{\mathcal E}$ excitatory (light orange) and $N_{\mathcal I}$ inhibitory (dark green) neurons. (b) Circle radius $r$ computed from Eq.~\eqref{eq_radius_ev} 
for the random connectivity scheme. Lines of equal radius ($r = 1$) are plotted in the
$(p, c_{\mathcal E})$-plane for $\bar{c} = 1,0,-1$ (solid green, dashed violet, dotted orange lines).
Parameters: $N_{\mathcal E}=400$, $N_{\mathcal I}=100$. 
(c) Largest absolute value $r$ of all transverse eigenvalues, averaged
over 100 sample matrices each representing a Mexican hat (left) or inverse Mexican hat (right) spatial map connectivity scheme: Excitatory and inhibitory neurons lie uniformly distributed on a two-dimensional sheet with unity edge length and periodic boundary conditions. The probability densities for excitatory and inhibitory connections are normalized Gaussian functions of the spatial distance between neurons with standard deviations $\sigma_{\mathcal E}$ and $\sigma_{\mathcal I}$, respectively. Lines of equal absolute value ($r = 1$) are plotted with color code as in (b).
Parameters: $N_{\mathcal E}=400$, $N_{\mathcal I}=100$, $p=0.2$; $\sigma_{\mathcal E} = 0.07$, $\sigma_{\mathcal I} \in [0.07, 0.35]$ (left) and $\sigma_{\mathcal E} \in [0.07, 0.35]$, $\sigma_{\mathcal I} = 0.07$ (right). $c_{\mathcal I}$ in (b) and (c) via constant row sum condition.
\label{fig2}}
\end{figure*}
%
The MSF provides general information about the stability of the synchronous state for many different coupling matrices. We now evaluate the results shown in Fig.~\ref{fig1} for two qualitatively
different and biologically important classes of coupling matrices. For both classes, the homogeneous population of aEIF
neurons is split into two subpopulations which make excitatory ($c_{\mathcal E}>0$) and inhibitory ($c_{\mathcal I}<0$) connections only, cf. Fig.~\ref{fig2}a. For one class of coupling matrices (random) the connections are otherwise random, i.e., connections between any two neurons in
the network are chosen with equal probability. For the other class of coupling matrices (spatial maps) neurons lie on a two-dimensional sheet and the connection probability between two neurons depends on the spatial distance between them. 

Figure~\ref{fig2}b shows the results of the eigenvalue spectrum for the random connectivity scheme.
One eigenvalue of the connection matrix is real and equal to the row sum $\bar{c}$. The other eigenvalues lie within a circle with radius
\begin{equation}
r = \sqrt{p (1-p) \left( N_{\mathcal E}c_{\mathcal E}^2 + N_{\mathcal I}c_{\mathcal I}^2 \right)} 
\label{eq_radius_ev}
\end{equation}
centered at the origin of the eigenvalue plane \cite{Gray2009b, Rajan2006}. $p$ is the connection
probability, $N_{\mathcal E}$, $N_{\mathcal I}$ are the number of neurons in the two different
subpopulations and $c_{\mathcal E} >0$, $c_{\mathcal I} <0$ are the excitatory and inhibitory coupling strengths, respectively. The curves separate the parameter regimes with $r<1$ and $r>1$. Figure~\ref{fig2}c shows the corresponding results for the spatial map scheme, which were calculated
numerically for both Mexican hat (range of excitatory couplings on average smaller than range of
inhibitory couplings) and inverse Mexcian hat (range of excitatory couplings on
average larger than range of inhibitory couplings) connection matrices. 
The maximum size of the eigenvalues depends on the absolute values of the coupling strengths. If
they are not too strong, the eigenvalues of all three types of coupling matrices lie within the unit
disc in the eigenvalue plane (highlighted in Fig.~\ref{fig1}). We conclude that if synchrony is
stable in a random network with sparse couplings, it is also stable in a spatially structured
network with local couplings only. This has interesting implications for network models of cortical
areas, because it implies that the stability of synchrony in a model of a local cortical column can predict
the stability of global synchronous activity in a model of a spatially structured cortical map and vice versa.

\section{V. Master Stability Function for cluster states}
Complete synchronization in a large network is a special phenomenon.
Often cluster states emerge, for which the network splits into subgroups of
elements that show isochronous synchrony internally, but there may be a phase lag or different dynamics between them \cite{Choe2010, SEL12, WIL13}. The stability of cluster states can
also be analyzed using the MSF
formalism \cite{SOR07, DAH12}, which we now formulate for
non-smooth dynamical systems.

\begin{figure*}
  \includegraphics[width=0.95\linewidth]{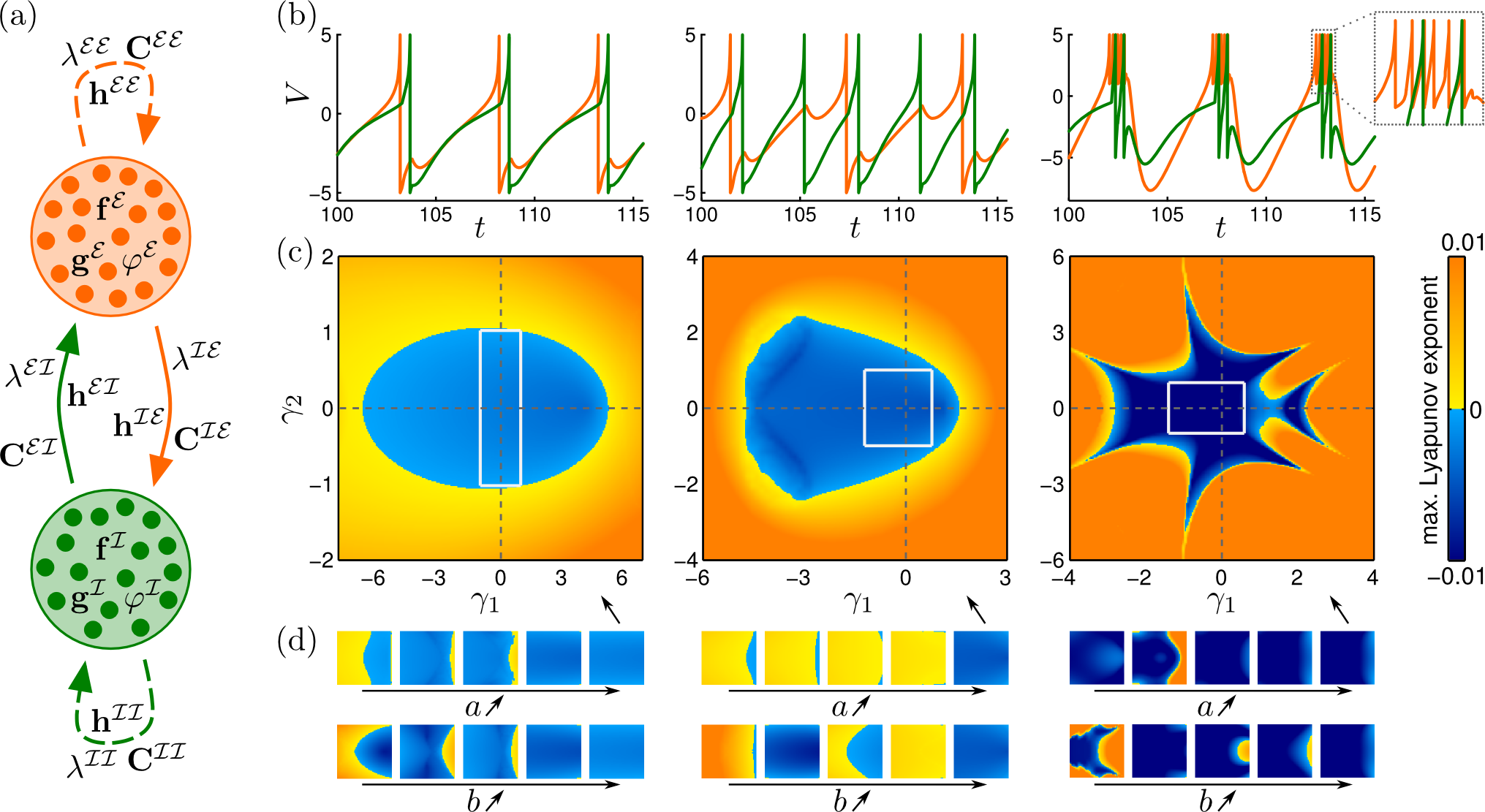} 
  \caption{(Color online) Stability of cluster states in networks with 
    two groups of excitatory, $\mathcal{E}$, and inhibitory, $\mathcal{I}$, neurons which differ in
   their local dynamics.
 	(a) Scheme of the network, 
    according to Eqs.~\eqref{eq:1}--\eqref{eq:2}. 
    (b) Evolution of the membrane voltage $V$ of synchronized excitatory
    (light orange) and inhibitory neurons (dark green) in the network
		for three different parameter sets:
    excitatory neurons: $V_r = -5$, $I = 3.725$ (left and center), 
    $V_r = 2$, $I = 25$ (right); inhibitory neurons: $I = 0$ (left and right), 
    $I = 1.5$ (center). 
    $\lambda^{\mathcal{E}\mathcal{E}}=\lambda^{\mathcal{I}\mathcal{E}}=5$,
    $\lambda^{\mathcal{E}\mathcal{I}}=\lambda^{\mathcal{I}\mathcal{I}}=-5$ (left and center);
    $\lambda^{\mathcal{E}\mathcal{E}}=\lambda^{\mathcal{I}\mathcal{E}}=25$,
    $\lambda^{\mathcal{E}\mathcal{I}}=\lambda^{\mathcal{I}\mathcal{I}}=-25$ (right).
    Other parameters: $a = 0.5$, $b = 2.5$, 
    $\tau_w = 10$, $\tau_s = 0.2$, $V_{th} = 5$ for the excitatory, and
    $a = 0.1$, $b = 0.25$, 
    $\tau_w = 10$, $\tau_s = 0.5$, $V_r = -5$, $V_{th} = 5$ for the inhibitory neurons; $\tau = 0.1$. (c) MSF for each of the three scenarios in (b). The maximum Lyapunov
    exponent is shown as a function of the real eigenvalues $\gamma_1, \gamma_2$. White rectangles indicate the unit square in the eigenvalue plane. (d) MSF of the unit square as in (c), but with $a = 0.1, 0.2, 0.3, 0.4, 0.5$ (top, left to right) and $b = 0.5, 1, 1.5, 2, 2.5$ (bottom) for the excitatory neurons. 
  \label{fig3}}
\end{figure*}
Consider a network which consists of two groups ($\mathcal E$,
$\mathcal I$) of elements which
may differ in their local dynamics (Fig.~\ref{fig3}a). The connections of elements within and between the
groups are given by the four coupling matrices $\mathbf{C}^{\mathcal{E}\mathcal{E}}$,
$\mathbf{C}^{\mathcal{I}\mathcal{I}}$, $\mathbf{C}^{\mathcal{E}\mathcal{I}}$, and
$\mathbf{C}^{\mathcal{I}\mathcal{E}}$, for which we assume unity row sum.
The dynamics of the network is given by:
\begin{align}
  \dot{\mathbf{x}}_i^{k\hphantom{,+}} &\!=\mathbf{f}^{k}(\mathbf{x}_i^{k}) 
  &&\!\!\!\!\!\!\!+\!
  \lambda^{kk}\sum_{j=1}^{N_k} c_{ij}^{kk}
  \mathbf{h}^{kk}(\mathbf{x}_{j,\tau}^{k}) && \nonumber\\
  &&&\!\!\!\!\!\!\!+\!
  \lambda^{kl}\sum_{j=1}^{N_l} c_{ij}^{kl} \mathbf{h}^{kl}(\mathbf{x}_{j,\tau}^{l})
  &&\!\!\!\!\!\text{if} \: \varphi^{k}(\mathbf{x}_i^{k}) \!\neq\! 0,  \label{eq:1} \\
  \mathbf{x}_i^{k,+} &\!= \mathbf{g}^{k}(\mathbf{x}_i^{k}) &&&&\!\!\!\!\!\text{if} \: \varphi^{k}(\mathbf{x}_i^{k})\!=\! 0, \label{eq:2}
\end{align}
where $k,l \in \{\mathcal{E},\mathcal{I}\}$ and $k \neq l$. The coupling strengths of the intra- and inter-group connections are scaled by factors $\lambda^{kk}, \lambda^{kl} \in
\mathbb{R}$, respectively. In order to simplify notation, the propagation delay $\tau$ is assumed constant.

In a cluster state, all elements which belong to a group of the network are synchronized and follow the same trajectory $\mathbf{x}_s^{k}\equiv \mathbf{x}_i^{k}$, $i = 1,\dots,N_k$. Trajectories of elements which belong to different groups, however, may be different. The variational equations for the perturbations from the synchronous solution, i.e., the master stability equations, are
then given by (cf. Eq.~\eqref{eq_mse1} for the homogenous case):
\begin{align}
\dot{\boldsymbol{\zeta}}^{k\hphantom{,+}} \! &\!\!\!= &&\!\!\!\!\!\!\!\!\!\!\!\!\!\!\!\!D
\mathbf{f}^{k}(\mathbf{x}_s^{k}) \boldsymbol{\zeta}^{k}
+ \lambda^{kk}\gamma_{1} D \mathbf{h}^{kk}(\mathbf{x}_{s,\tau}^{k}) \boldsymbol{\zeta}_{\tau}^{k}
 \nonumber\\
&&&\!\!\!\!\!\!\!\!\!\!\!\!\!\!\!\!+ \lambda^{kl}\gamma_{2} D \mathbf{h}^{kl}(\mathbf{x}_{s,\tau}^{l}) \boldsymbol{\zeta}_{\tau}^{l}, \label{eq:14} 
\end{align}
for $t, t \! - \! \tau \notin \lbrace t_s^{k} \rbrace $ and $t \! - \! \tau
\notin \lbrace t_s^{l} \rbrace $. $\gamma_{1}$ and $\gamma_{2}$ are the
eigenvalues of the block matrices $\mathbf{Q}_{1}=\left(\begin{smallmatrix}
  \mathbf{C}^{\mathcal{E}\mathcal{E}} & \mathbf{0} \\
  \mathbf{0} & \mathbf{C}^{\mathcal{I}\mathcal{I}}
\end{smallmatrix}\right)$ and $\mathbf{Q}_{2}=\left(\begin{smallmatrix}
  \mathbf{0} & \mathbf{C}^{\mathcal{E}\mathcal{I}} \\
  \mathbf{C}^{\mathcal{I}\mathcal{E}} & \mathbf{0}
\end{smallmatrix}\right)$, respectively \cite{SOR07,DAH12}.
The variational equations - one for each group - must again be
complemented by the linearized transition conditions (cf. Eqs.~\eqref{eq_mse2}--\eqref{eq_mse3} for the homogeneous case):
\begin{align}
\boldsymbol{\zeta}^{k,+} &\!= \mathbf{A}^{k} \boldsymbol{\zeta}^{k} &&&&\!\!\!\!\!\!\!\!\!\!\!\!\!\!\!\!\!\!\!\!\!t \in \lbrace t_s^{k} \rbrace , \label{eq:15} \\
\boldsymbol{\zeta}^{k,+} &\!= \boldsymbol{\zeta}^{k} + \lambda^{kk}\gamma_{1} \mathbf{B}^{k} \boldsymbol{\zeta}_{\tau}^{k} &&&&\!\!\!\!\!\!\!\!\!\!\!\!\!\!\!\!\!\!\!\!\!t \! - \! \tau \in \lbrace t_s^{k} \rbrace , \label{eq:16}\\
\boldsymbol{\zeta}^{l,+} &\!= \boldsymbol{\zeta}^{l} + \lambda^{lk}\gamma_{2}\mathbf{B}^{k} \boldsymbol{\zeta}_{\tau}^{k} &&&&\!\!\!\!\!\!\!\!\!\!\!\!\!\!\!\!\!\!\!\!\!t \! - \! \tau \in \lbrace t_s^{k} \rbrace . \label{eq:17}
\end{align}
$\mathbf{A}^{k}$ and $\mathbf{B}^{k}$ only depend on $\mathbf{x}_s^{k}$ just before and after the
discontinuity, see Appendix II. 
However, we have to require that the matrices $\mathbf{Q}_{1}$ and
$\mathbf{Q}_{2}$ commute, because only then a common set of eigenvectors exists and the pairs $(\gamma_1, \gamma_2)$ of eigenvalues are well defined~\cite{DAH12}. In order to assess the
stability of a particular
cluster state, the MSF is then evaluated for those pairs. 

\section{VI. Stability of cluster states for networks of adaptive neurons}
We apply our MSF formalism to a network of aEIF neurons, where the
excitatory and the inhibitory subpopulations now have different local dynamics (cf. Fig.~\ref{fig3}a).
The aEIF neurons of the inhibitory subpopulation are weakly adaptive, and parameters are chosen such that they mimick the dynamics of the fast spiking inhibitory interneurons in the neocortex~\cite{Destexhe2009}. The aEIF neurons of the excitatory subpopulation are strongly adaptive,
and parameters are chosen such that they either mimick the dynamics of the pyramidal neurons in the
neocortex (Fig.~\ref{fig3}b,c, left and center) or the dynamics of intrinsically bursting neurons
(Fig.~\ref{fig3}b,c, right), which - when activated - produce repeated events of rapid spiking. Bursting dynamics is mediated by spike-based adaptation for neurons with an increased reset membrane potential \cite{Naud2008}.
Figure~\ref{fig3}c shows the MSF for real eigenvalues $\gamma_1, \gamma_2$ of the intra- and inter-group coupling matrices. If connections are symmetric ($c_{ij}^{kl}=c_{ji}^{kl}$, $k,l \in \{\mathcal{E},\mathcal{I}\}$) and no autapses ($c_{ii}^{kk}=0$) exist, the eigenvalues $\gamma_1, \gamma_2$ are real and lie
within the unit square of the eigenvalue plane \footnote{This follows from Gershgorin's circle theorem, because
the coupling matrices $\mathbf{C}^{kl}$ ($k,l \in \{\mathcal{E},\mathcal{I}\}$) all have non-negative entries with unity row sum. Excitatory and
inhibitory interactions are generated via the overall coupling parameters $\lambda^{k\mathcal{E}} >
0$ and $\lambda^{k\mathcal{I}} < 0$}.
The corresponding Lyapunov exponents 
are all negative (Fig.~\ref{fig3}c). However, they can become positive when subthreshold or spike-based adaptation for the excitatory neurons is decreased (Fig.~\ref{fig3}d), indicating a potential loss of stability for the particular cluster states. We conclude that the stability of many qualitatively different cluster states is controlled by neuronal adaptation: 
Oscillatory activity of excitatory and inhibitory neurons with a phase lag between the
two groups (Fig.~\ref{fig3}b,c, left), inhibitory neurons spiking repetitively at a higher rate than
excitatory neurons (Fig.~\ref{fig3}b,c, center), and (3) bursting behavior (Fig.~\ref{fig3}b,c, right).
These results hold for qualitatively different classes of coupling matrices: Networks with constant all-to-all intra- and random (but symmetric, $c_{ij}^{kl} = c_{ji}^{lk}$) inter-group connections, for example,
show the same synchronization behavior as one-dimensional spatial networks with local (symmetric) connectivity ($N_\mathcal{E} = N_\mathcal{I}$).

\section{VII. Conclusion}
In summary, we have presented an MSF formalism to study synchrony and cluster states
in networks of non-smooth dynamical systems. Using our formalism we have shown that neuronal adaptation controls the stability of such network states 
for many qualitatively different classes of coupling matrices. Hence, our approach provides a powerful method to analyze biologically relevant dynamical states of typical cortical network models in a general setting. For example, it could be used to examine the stability of synchronous and clustered network states under influence of structural plasticity (i.e., activity-dependent changes of coupling strengths) \cite{Cateau2008,Lubenov2008}, irrespective of a specific coupling topology. 
Another interesting application of our approach in neuroscience is pulsed brain stimulation, such as transcranial magnetic stimulation \cite{Hallett2000} or deep brain stimulation \cite{Kringelbach2007}, where our method could provide general results, 
e.g., when using models of coupled neuronal populations and pulse-like forcing \cite{Esser2005,Tass2003}.
Our formalism is, however, not limited to the neuronal network level, but may be applied to a variety of diverse networks, e.g., coupled eukaryotic model cells \cite{Battogtokh2006a}, hybrid models of genetic networks \cite{Perkins2010,Zhang2011}, electronic circuits such as networks of Chua's circuits \cite{DeLellis2008}, hybrid switching systems in production technology \cite{AMA03}, or delay-coupled semiconductor lasers with challenging perspectives towards reservoir computing \cite{SOR13}. Promising applications are also anticipated for networks of interacting oscillators in the context of cognitive processing \cite{Deco2011}. Here, the MSF can provide valuable information about the robustness of synchrony and locking dynamics against perturbations of the brain's connectivity patterns.

\section{Acknowledgment}
This work was supported by DFG in the framework of SFB 910.

\setcounter{equation}{0}
\renewcommand{\theequation}{A.\arabic{equation}}

\section{Appendix I: Variational equation for full synchrony}

\noindent Here we derive the transition conditions (Eqs.~(6)--(9) of the paper) which complete the variational equation (5), i.e., the linearization of the network system Eqs.~(3)--(4) around the synchronous solution. The derivation builds upon previous work on dynamical systems with discontinuities~\cite{Mueller1995} and extends the case of uncoupled systems \cite{Ladenbauer2012}. \\ \\
Let $\underline{\mathbf{x}}_s(t)$ with $\underline{\mathbf{x}}_s \equiv (\mathbf{x}_s,\dots,\mathbf{x}_s)^T \!\!:  \mathbb{R}\!\rightarrow \! \mathbb{R}^{Nm}$ denote the synchronous solution of Eqs.~(3)--(4), i.e., the components $\mathbf{x}_s$ solve $\dot{\mathbf{x}} = \mathbf{f}(\mathbf{x}) + \bar{c} \, \mathbf{h}(\mathbf{x}_{\tau})$.
Furthermore, let $\underline{\boldsymbol{\xi}}$ be the solution of the linearized system Eq.~(5) for small initial perturbations, $\underline{\boldsymbol{\xi}} \equiv \underline{\boldsymbol{\xi}}_0 \!: [t_0 - \tau, t_0]\!\rightarrow \! \mathbb{R}^{Nm}$, 
$|| \underline{\boldsymbol{\xi}}_0(s) || < \epsilon$, and let $t_s$ be the first time point $>t_0$ at which $\varphi(\mathbf{x}_s) = 0$.
%
%
We define $\underline{\tilde{\mathbf{x}}} \equiv \underline{\mathbf{x}}_s + \underline{\boldsymbol{\xi}}$ and denote by $t_i$ the time point (closest to $t_s$) at which $\varphi(\tilde{\mathbf{x}}_i) = 0$. 

\begin{figure*}[!htp]
\includegraphics[width=0.75\linewidth]{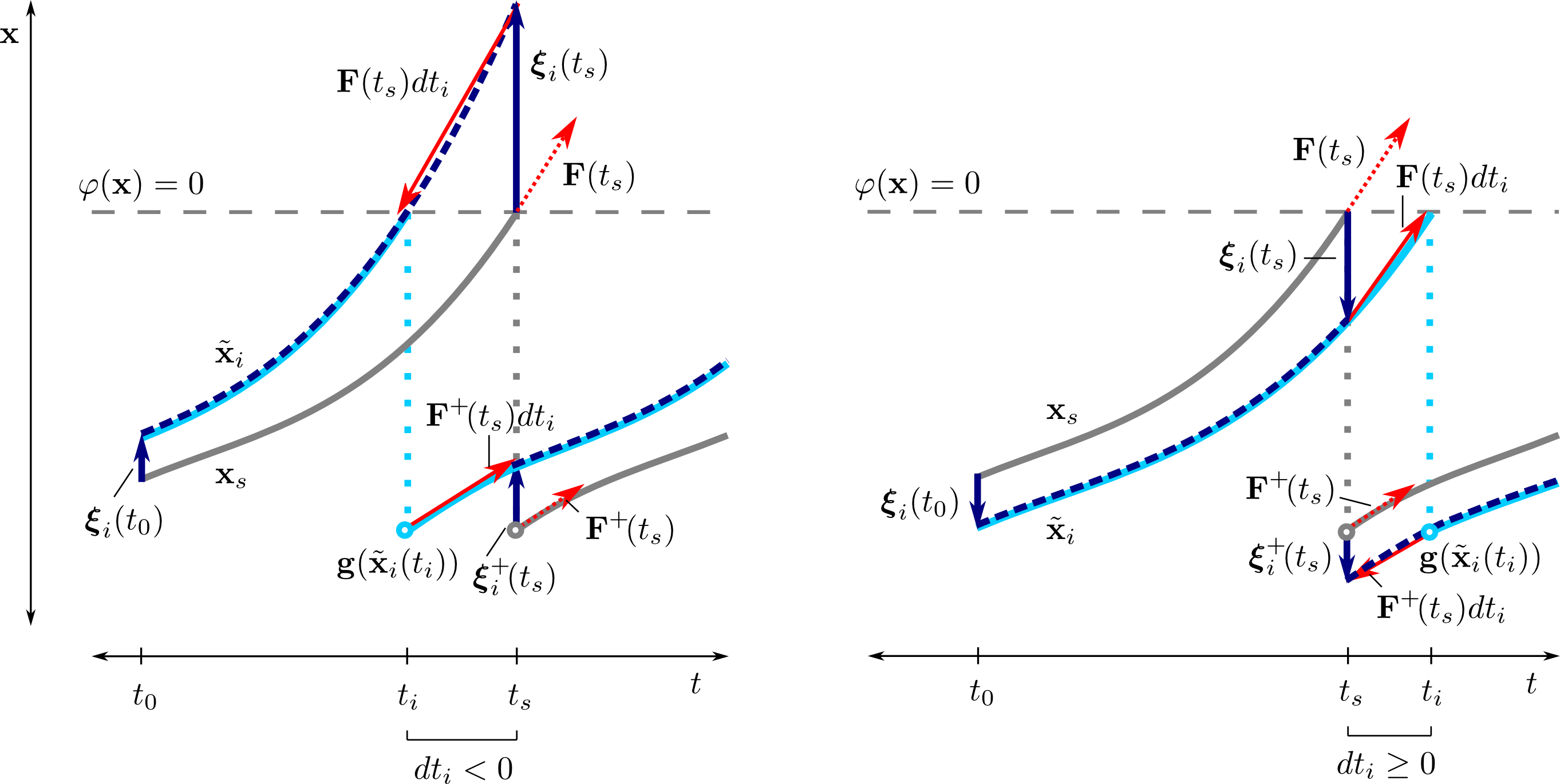}
\caption{(Color online) Approximation scheme for the perturbation $\boldsymbol{\xi}_i^+(t_s)$ just after the discontinuity of $\mathbf{x}_s$, illustrated for scalar-valued functions $\mathbf{x}_i$ ($m=1$). Evolution of $\tilde{\mathbf{x}}_i \equiv \mathbf{x}_s + \boldsymbol{\xi}_i$ (dashed dark blue lines), where $\mathbf{x}_s$ (solid grey lines) is a component of the synchronous solution of the system Eqs.~(3)--(4) and $\boldsymbol{\xi}_i$ is a solution component of Eq.~(5) with initial condition $\underline{\boldsymbol{\xi}} \equiv \underline{\boldsymbol{\xi}}_0$, 
for $dt_i <0$ (left) and $dt_i >0$ (right). Note that $\tilde{\mathbf{x}}_i$ is not discontinuous at $t_i \neq t_s$, because it is not a solution component of Eqs.~(3)--(4). The $i$-th solution component of Eqs.~(3)--(4) with initial trajectory $\underline{\mathbf{x}}_s + \underline{\boldsymbol{\xi}}_0$ is indicated (solid light blue lines). Dashed grey lines mark the set $\{\mathbf{x} \,|\, \varphi(\mathbf{x}) = 0\}$ which constitutes the threshold hyperplane here. If a solution $\mathbf{x}$ of Eq.~(3) reaches this hyperplane from below, a discontinuity occurs: $\mathbf{x}^+ = \mathbf{g}(\mathbf{x})$, cf. Eq.~(4). Solid red arrows indicate the vectors $\mathbf{F}(t_s) dt_i$ and $\mathbf{F}^+(t_s) dt_i$ used in the approximation of $\boldsymbol{\xi}_i^+(t_s)$. The vectors $\mathbf{F}(t_s)$ and $\mathbf{F}^+(t_s)$ are indicated by dotted red arrows.
\label{supfig1}}
\end{figure*}

\subsection{A. Transition condition at the discontinuities}

\noindent In the following we provide the linear relationship between $\underline{\boldsymbol{\xi}}(t_s)$ and $\underline{\boldsymbol{\xi}}^+(t_s)$, the perturbations just before and after the discontinuity in $\underline{\mathbf{x}}_s$. 
Starting from $\tilde{\mathbf{x}}_i(t_s)$ 
we estimate $\tilde{\mathbf{x}}_i(t_i)$, apply the discontinuous transition Eq.~(4) and then approximate $\tilde{\mathbf{x}}_i^+(t_s)$, as illustrated in Fig.~\ref{supfig1}:
\begin{align}
\tilde{\mathbf{x}}_i^+(t_s) &\approx \mathbf{g}(\mathbf{x}_s(t_s) + \boldsymbol{\xi}_i(t_s) + \tilde{\mathbf{F}}_i(t_s) dt_i) \nonumber \\
& \quad - \tilde{\mathbf{F}}_i^+(t_i) dt_i, \label{eq_pert_ts1} \\
&\approx \mathbf{x}_s^+(t_s) + D \mathbf{g}(\mathbf{x}_s(t_s)) 
(\boldsymbol{\xi}_i(t_s) + \tilde{\mathbf{F}}_i(t_s) dt_i) \nonumber \\
& \quad - \tilde{\mathbf{F}}_i^+(t_i) dt_i, \label{eq_pert_ts2}
\end{align}
where we have used the definitions $dt_i \equiv t_i - t_s$, 
\begin{align}
\tilde{\mathbf{F}}_i(t_s) &\equiv \mathbf{f}(\tilde{\mathbf{x}}_i(t_s)) + \sum_{j=1}^{N} c_{ij} \mathbf{h}(\tilde{\mathbf{x}}_j(t_s - \tau)), \label{eq_F_def1} \\
\tilde{\mathbf{F}}_i^+(t_i) &\equiv \mathbf{f}(\tilde{\mathbf{x}}_i^+(t_i)) + \sum_{j=1}^{N} c_{ij} \mathbf{h}(\tilde{\mathbf{x}}_j^+(t_i - \tau)), \label{eq_F_def4}
\end{align}
%
%
and Taylor expansion. Since $dt_i$, $\tilde{\mathbf{F}}_i(t_s)$ and $\tilde{\mathbf{F}}_i^+(t_i)$ are unknown we use the following approximations. First we expand $\tilde{\mathbf{F}}_i(t_s)$, 
\begin{align}
\tilde{\mathbf{F}}_i(t_s) &\approx \mathbf{f}(\mathbf{x}_s(t_s)) + \bar{c} \, \mathbf{h}(\mathbf{x}_s(t_s - \tau)) \equiv \mathbf{F}(t_s). \label{eq_F_def3}  
\end{align}  
\\
In order to describe $dt_i$ we approximate $\tilde{\mathbf{x}}_i(t_i)$ 
as
\begin{align}
\tilde{\mathbf{x}}_i(t_i) &\approx \tilde{\mathbf{x}}_i(t_s) + \tilde{\mathbf{F}}_i(t_s) dt_i \\
&\approx \mathbf{x}_s(t_s) + \boldsymbol{\xi}_i(t_s) + \mathbf{F}(t_s) dt_i 
\end{align}
and apply $\varphi$ on both sides, which leads to
\begin{align}
0 &\approx \varphi(\mathbf{x}_s(t_s) + \boldsymbol{\xi}_i(t_s) + \mathbf{F}(t_s) dt_i)  
\label{eq_dti1} \\
&\approx D \varphi(\mathbf{x}_s(t_s)) (\boldsymbol{\xi}_i(t_s) + \mathbf{F}(t_s) dt_i)
\end{align}
We obtain a first order estimation for $dt_i$, 
%
\begin{align}
dt_i &\approx - \frac{D \varphi(\mathbf{x}_s(t_s)) \boldsymbol{\xi}_i(t_s)}{D \varphi(\mathbf{x}_s(t_s)) \mathbf{F}(t_s)}. \label{eq_dti2}
\end{align}
%
%
Next we expand $\tilde{\mathbf{F}}_i^+(t_i)$, 
\begin{align}
\tilde{\mathbf{F}}_i^+(t_i) &= \mathbf{f}(\mathbf{g}(\tilde{\mathbf{x}}_i(t_i))) + \sum_{j=1}^{N} c_{ij} \mathbf{h}(\tilde{\mathbf{x}}_j^+(t_i - \tau)) \label{eq_F_def5} \\
&\approx \mathbf{f}(\mathbf{g}(\tilde{\mathbf{x}}_i(t_s))) + \sum_{j=1}^{N} c_{ij} \mathbf{h}(\tilde{\mathbf{x}}_j^+(t_s - \tau)) \label{eq_F_def6} \\
&\approx \mathbf{f}(\mathbf{x}_s^+(t_s)) + \bar{c} \, \mathbf{h}(\mathbf{x}_s^+(t_s - \tau)) \equiv \mathbf{F}^+(t_s). \label{eq_F_def7}
\end{align}
%
We substitute $dt_i$, $\tilde{\mathbf{F}}_i(t_s)$, $\tilde{\mathbf{F}}_i^+(t_i)$ in Eq.~\eqref{eq_pert_ts2} and subtract $\mathbf{x}_s^+(t_s)$ on both sides to obtain the first order approximation
%
%
\begin{align}
\boldsymbol{\xi}_i^+ &\approx \mathbf{A} \boldsymbol{\xi}_i(t_s) 
\label{eq_transition_cond1} \\ 
\mathbf{A} &\equiv
D\mathbf{g}(\mathbf{x}_s(t_s)) \nonumber \\ 
& \quad + \frac{\left[ \mathbf{F}^+(t_s) - D \mathbf{g}(\mathbf{x}_s(t_s)) \mathbf{F}(t_s) \right] D \varphi(\mathbf{x}_s(t_s))}{D \varphi(\mathbf{x}_s(t_s)) \mathbf{F}(t_s)} \label{eq_transition_cond2}
\end{align}
%
The transition Eqs.~\eqref{eq_transition_cond1}--\eqref{eq_transition_cond2} holds at all times $t_s \in \lbrace t_s \rbrace$ at which $\mathbf{x}_s$ changes discontinuously.

\subsection{B. Transition condition at the kinks}

\begin{figure}[!htp]
\includegraphics[width=0.85\linewidth]{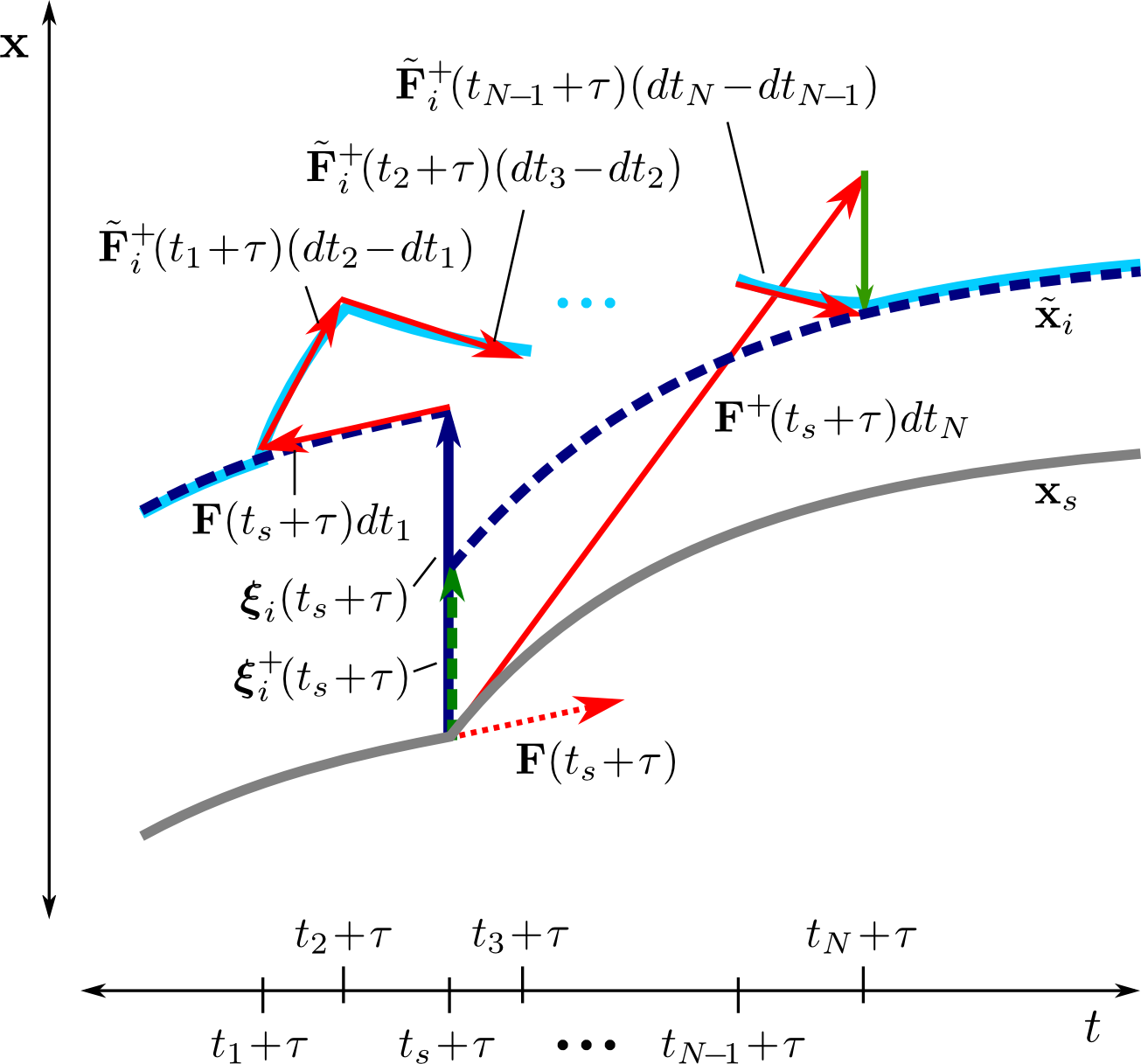}
\caption{(Color online) Approximation scheme for the perturbation $\boldsymbol{\xi}_i^+(t_s+\tau)$ just after the kink in $\mathbf{x}_s$, illustrated for scalar-valued functions $\mathbf{x}_i$ ($m=1$). Evolution of the synchronous solution component $\mathbf{x}_s$ (solid grey line), $\tilde{\mathbf{x}}_i \equiv \mathbf{x}_s + \boldsymbol{\xi}_i$ (dashed dark blue lines) and the $i$-th solution component of Eqs.~(3)--(4) with initial trajectory $\underline{\mathbf{x}}_s + \underline{\boldsymbol{\xi}}_0$ (solid light blue lines). Solid red arrows indicate vectors that are used in the approximation, as explained in the text. The solid green arrow represents the difference between $\boldsymbol{\xi}_i^+(t_s+\tau)$ (dashed green arrow) and $\boldsymbol{\xi}_i(t_s+\tau)$ (solid dark blue arrow).
\label{supfig2}}
\end{figure}

\noindent Next we derive the linear relationship between $\underline{\boldsymbol{\xi}}(t_s+\tau)$ and $\underline{\boldsymbol{\xi}}^+(t_s+\tau)$, the perturbations just before and after the kink in $\underline{\mathbf{x}}_s$ caused by the delayed coupling. 
We assume w.l.o.g. that the time points $t_i$ are ordered $t_1 \leq t_2 \leq \dots \leq t_N$. Starting from $\tilde{\mathbf{x}}_i(t_s+\tau)$ we estimate first $\tilde{\mathbf{x}}_i^+(t_1+\tau)$, then iteratively $\tilde{\mathbf{x}}_i^+(t_2+\tau), \dots, \tilde{\mathbf{x}}_i^+(t_N+\tau)$ and finally $\tilde{\mathbf{x}}_i^+(t_s+\tau)$, see Fig.~\ref{supfig2} for an illustration:  
\begin{align}
\tilde{\mathbf{x}}_i^+(t_1+\tau) &\approx \tilde{\mathbf{x}}_i(t_s+\tau) + 
\tilde{\mathbf{F}}_i(t_s+\tau) dt_1 \label{eq_coupling_step1a} \\
&\approx \tilde{\mathbf{x}}_i(t_s+\tau) + \mathbf{F}(t_s+\tau) dt_1 , \label{eq_coupling_step1b}
\end{align}
where we have applied Eqs.~\eqref{eq_F_def1} and \eqref{eq_F_def3}. Using the definition~\eqref{eq_F_def4} we approximate iteratively (to first order)
\begin{align}
\tilde{\mathbf{x}}_i^+(t_{k+1}+\tau) &\approx \tilde{\mathbf{x}}_i(t_k+\tau) \nonumber \\
&\quad + \tilde{\mathbf{F}}_i^+(t_k+\tau) (dt_{k+1} - dt_k) \label{eq_coupling_step2}
\end{align}
for $k=1, \dots ,N-1$. 
Since we do not know $\tilde{\mathbf{F}}_i^+(t_k+\tau)$, we approximate it as 
\begin{align}
\tilde{\mathbf{F}}_i^+(t_k+\tau) &\approx \mathbf{f}(\mathbf{x}_s(t_s+\tau)) + \sum_{j=1}^{k} c_{ij} \mathbf{h}(\mathbf{x}_s^+(t_s))  \nonumber \\
& \quad + \sum_{j=k+1}^{N} c_{ij} \mathbf{h}(\mathbf{x}_s(t_s)), \label{eq_coupling_step2_detail}
\end{align}
where we have used that 
\begin{equation}
\mathbf{h}(\tilde{\mathbf{x}}_j(t_k)) \approx \left\{
\begin{array}{l l} 
 \mathbf{h}(\mathbf{x}_s^+(t_s)) & \quad \text{for $k \geq j$} \\
 \mathbf{h}(\mathbf{x}_s(t_s)) & \quad \text{for $k < j$}, \\
\end{array} \right.
\end{equation}
and that $\mathbf{x}_s$ is continuous at $t_s + \tau$, assuming $t_s + \tau \notin \lbrace t_s \rbrace$. Using Eq.~\eqref{eq_coupling_step2} we can now estimate $\tilde{\mathbf{x}}_i^+(t_N+\tau)$
\begin{align}
\tilde{\mathbf{x}}_i^+(t_N+\tau) &\approx \tilde{\mathbf{x}}_i(t_1+\tau) + 
\sum_{k=1}^{N-1} \tilde{\mathbf{F}}_i^+(t_k+\tau) (dt_{k+1} - dt_k) \nonumber \\
&\approx \tilde{\mathbf{x}}_i(t_s+\tau) + \mathbf{F}^+(t_s+\tau) dt_N \nonumber \\
& \quad \: + \left[ \mathbf{h}(\mathbf{x}_s(t_s)) - 
\mathbf{h}(\mathbf{x}_s^+(t_s)) \right] \sum_{j=1}^{N} c_{ij} dt_j , \label{eq_coupling_step3}
\end{align}
where in Eq.~\eqref{eq_coupling_step3} we have applied Eqs.~\eqref{eq_coupling_step1b},~\eqref{eq_coupling_step2_detail}, the definitions~\eqref{eq_F_def3},~\eqref{eq_F_def7} and telescopic cancelling. 
We use the fact that $\tilde{\mathbf{x}}_i^+(t_N+\tau)$ can also be expressed as
\begin{align}
\tilde{\mathbf{x}}_i^+(t_N+\tau) &\approx \tilde{\mathbf{x}}_i^+(t_s+\tau) + \mathbf{F}^+(t_s+\tau) dt_N , \label{eq_coupling_step3a}
\end{align}
and subtract $\mathbf{F}^+(t_s+\tau) dt_N$ and $\mathbf{x}_s(t_s+\tau)$ on both sides of Eq.~\eqref{eq_coupling_step3}, which yields
\begin{align}
\boldsymbol{\xi}_i^+(t_s+\tau) &\approx \boldsymbol{\xi}_i(t_s+\tau) + 
\left[ \mathbf{h}(\mathbf{x}_s(t_s)) - \mathbf{h}(\mathbf{x}_s^+(t_s)) \right] \nonumber \\
& \quad \: \cdot \sum_{j=1}^{N} c_{ij} dt_j.
\label{eq_coupling_step4}
\end{align}
Note that Eqs.~\eqref{eq_coupling_step3} and \eqref{eq_coupling_step3a} are first order approximations.
Finally we substitute $dt_j$ in Eq.~\eqref{eq_coupling_step4} using Eq.~\eqref{eq_dti2} to obtain
\begin{align}
\underline{\boldsymbol{\xi}}^+(t_s+\tau) &\approx \underline{\boldsymbol{\xi}}(t_s+\tau) + 
\left[ \mathbf{I}_N \otimes \frac{\mathbf{h}(\mathbf{x}_s^+(t_s)) - \mathbf{h}(\mathbf{x}_s(t_s))}{D \varphi(\mathbf{x}_s(t_s)) \mathbf{F}(t_s)} \right] \nonumber \\
& \quad \: \cdot  \left[ \mathbf{C} \otimes D \varphi(\mathbf{x}_s(t_s)) \right] \underline{\boldsymbol{\xi}}(t_s)  \\
&=  \underline{\boldsymbol{\xi}}(t_s+\tau) + \left[ \mathbf{C} \otimes  \mathbf{B} \right] \underline{\boldsymbol{\xi}}(t_s), \label{eq_coupling_trans_cond} \\
\mathbf{B} &\equiv \frac{\left[ \mathbf{h}(\mathbf{x}_s^+(t_s)) - \mathbf{h}(\mathbf{x}_s(t_s)) \right] D \varphi(\mathbf{x}_s(t_s))}{D \varphi(\mathbf{x}_s(t_s)) \mathbf{f}(\mathbf{x}_s(t_s)) + \bar{c} \, \mathbf{h}(\mathbf{x}_s(t_s - \tau))}.
\end{align}
The transition Eq.~\eqref{eq_coupling_trans_cond} holds at all times $t_s+\tau$ for $ t_s \in \lbrace t_s \rbrace$.

It should be noted that the solution $\underline{\boldsymbol{\xi}}$ of the complete linearized system, Eq.~(5) with transition conditions \eqref{eq_transition_cond1}, \eqref{eq_transition_cond2}, \eqref{eq_coupling_trans_cond}, approximates the evolving true perturbation given by the solution of Eqs.~(3)--(4) with the small initial perturbation $\underline{\boldsymbol{\xi}} \equiv \underline{\boldsymbol{\xi}}_0$. 
The true perturbation is approximated to first order at all times except for small time intervals around $t_s$, $t_s+\tau$ for $ t_s \in \lbrace t_s \rbrace$ (with duration of the same order as the magnitude of the perturbation). During these short intervals the true perturbation can be large in magnitude since generally $t_i \neq t_s$. The transition conditions guarantee that immediately after each of these intervals (which contain the instants of discontinuities and kinks) the true perturbation is approximated to first order again.

\section{Appendix II: Variational equation for cluster states} \label{msf_cluster}

\noindent The transition conditions which complete the variational equation for synchronized groups of elements, i.e., the linearization of the system Eqs.~(14)--(15) around the cluster state, are derived analogously to the previous section. This analogy is demonstrated by the results presented below. \\ \\
%
The variational equation, which governs the dynamics of the perturbations 
$\underline{\boldsymbol{\xi}} \equiv (\boldsymbol{\xi}_1^{\mathcal{E}},\dots,\boldsymbol{\xi}_{N_{\mathcal{E}}}^{\mathcal{E}},\boldsymbol{\xi}_1^{\mathcal{I}},\dots,\boldsymbol{\xi}_{N_{\mathcal{I}}}^{\mathcal{I}})^T \!: \, \mathbb{R} \rightarrow \mathbb{R}^{(N_{\mathcal{E}}+N_{\mathcal{I}})m}$ is given by
\begin{align} \label{eq:3}
\dot{\underline{\boldsymbol{\xi}}} =&
\begin{pmatrix}
  \mathbf{I}_{N_\mathcal{E}} \otimes D
  \mathbf{f}^{\mathcal{E}}(\mathbf{x}_s^{\mathcal{E}}) & 0 \\
  0 & \mathbf{I}_{N_\mathcal{I}} \otimes D
  \mathbf{f}^{\mathcal{I}}(\mathbf{x}_s^{\mathcal{I}})
\end{pmatrix}
 \underline{\boldsymbol{\xi}}    \nonumber\\
 & +
 \begin{pmatrix}
   \mathbf{C}^{\mathcal{E}\mathcal{E}} \otimes D
   \mathbf{h}^{\mathcal{E}\mathcal{E}}(\mathbf{x}_{s,\tau}^{\mathcal{E}}) & \mathbf{C}^{\mathcal{E}\mathcal{I}} \otimes D
   \mathbf{h}^{\mathcal{E}\mathcal{I}} (\mathbf{x}_{s,\tau}^{\mathcal{I}}) \\
   \mathbf{C}^{\mathcal{I}\mathcal{E}} \otimes D
   \mathbf{h}^{\mathcal{I}\mathcal{E}}(\mathbf{x}_{s,\tau}^{\mathcal{E}}) &
   \mathbf{C}^{\mathcal{I}\mathcal{I}} \otimes D
   \mathbf{h}^{\mathcal{I}\mathcal{I}}(\mathbf{x}_{s,\tau}^{\mathcal{I}})
 \end{pmatrix}
 \underline{\boldsymbol{\xi}}_{\tau} 
\end{align}
for  $t, t \! - \! \tau \notin \lbrace t_s^{\mathcal{E}} \rbrace $ and
$t, t \! - \! \tau \notin \lbrace t_s^{\mathcal{I}} \rbrace $.
The transition conditions for the time points, at which the synchronized elements of each group 
change discontinuously, are expressed as
\begin{align} 
\underline{\boldsymbol{\xi}}^+ &= \begin{pmatrix}
  \mathbf{I}_{N_\mathcal{E}} \otimes \mathbf{A}^{\mathcal{E}} & 0 \\
  0 & 0
\end{pmatrix}
 \underline{\boldsymbol{\xi}} & &t \in \lbrace t_s^{\mathcal{E}} \rbrace , \label{eq:4} \\
\underline{\boldsymbol{\xi}}^+ &= \begin{pmatrix}
  0 & 0 \\
  0 & \mathbf{I}_{N_\mathcal{I}} \otimes \mathbf{A}^{\mathcal{I}}
\end{pmatrix}
 \underline{\boldsymbol{\xi}} & &t \in \lbrace t_s^{\mathcal{I}}
 \rbrace , \label{eq:5}
\end{align}
and the transition conditions for the time points at which the delay period has passed since these (discontinuous) events, read
\begin{align}
\underline{\boldsymbol{\xi}}^+ &= \underline{\boldsymbol{\xi}} +  \begin{pmatrix}
   \mathbf{C}^{\mathcal{E}\mathcal{E}} \otimes \mathbf{B}^{\mathcal{E}} & 0 \\
   \mathbf{C}^{\mathcal{I}\mathcal{E}} \otimes \mathbf{B}^{\mathcal{E}} & 0
 \end{pmatrix}
 \underline{\boldsymbol{\xi}}_{\tau} & &t \! - \! \tau \in \lbrace
 t_s^{\mathcal{E}} \rbrace \label{eq:6} \\
\underline{\boldsymbol{\xi}}^+ &= \underline{\boldsymbol{\xi}} +  \begin{pmatrix}
   0 & \mathbf{C}^{\mathcal{E}\mathcal{I}} \otimes \mathbf{B}^{\mathcal{I}} \\
   0 & \mathbf{C}^{\mathcal{I}\mathcal{I}} \otimes \mathbf{B}^{\mathcal{I}} 
 \end{pmatrix}
 \underline{\boldsymbol{\xi}}_{\tau} & &t \! - \! \tau \in \lbrace t_s^{\mathcal{I}} \rbrace .
\label{eq:7}
\end{align}
The matrices $\mathbf{A}^{k}$ and
$\mathbf{B}^{k}$ for
$k,l \in \lbrace \mathcal{E},\mathcal{I} \rbrace$, $ k \neq l$, 
in Eqs.~\eqref{eq:4}--\eqref{eq:7} are given by
\begin{widetext}
\begin{align} 
  \mathbf{A}^{k} &\equiv D \mathbf{g}^{k}(\mathbf{x}_s^{k}) + \frac{\left[ \mathbf{F}^{k}(\mathbf{x}_s^{k,+}, \mathbf{x}_{s,\tau}^{k}, \mathbf{x}_{s,\tau}^{l}) - D \mathbf{g}^{k}(\mathbf{x}_s^{k}) \mathbf{F}^{k}(\mathbf{x}_s^{k}, \mathbf{x}_{s,\tau}^{k}, \mathbf{x}_{s,\tau}^{l}) \right] D\varphi^{k}(\mathbf{x}_s^{k})}{D \varphi^{k}(\mathbf{x}_s^{k}) \mathbf{F}^{k}(\mathbf{x}_s^{k}, \mathbf{x}_{s,\tau}^{k}, \mathbf{x}_{s,\tau}^{l})} \label{eq:8} \\
  \mathbf{B}^{k} &\equiv \frac{\left[ \mathbf{F}^{k}(\mathbf{x}_s^{k},
      \mathbf{x}_{s,\tau +}^{k}, \mathbf{x}_{s,\tau}^{l}) - \mathbf{F}^{k}(\mathbf{x}_s^{k}, \mathbf{x}_{s,\tau}^{k}, \mathbf{x}_{s,\tau}^{l})
    \right] D \varphi^{k}(\mathbf{x}_{s,\tau}^{k})}{D
    \varphi^{k}(\mathbf{x}_{s,\tau}^{k}) \mathbf{F}^{k}(\mathbf{x}_{s,\tau}^{k},
    \mathbf{x}_{s,2 \tau}^{k}, \mathbf{x}_{s,2 \tau}^{l})}, \label{eq:9}
\end{align}  
\end{widetext}
where we have used the definition
\begin{align}
\mathbf{F}^{k}(\mathbf{x}^{k}, \mathbf{x}^{k}_{\tau},
\mathbf{x}^{l}_{\tau}) \equiv \mathbf{f}^{k}(\mathbf{x}^{k}) +
\lambda^{kk}\mathbf{h}(\mathbf{x}^{k}_{\tau}) +
\lambda^{kl}\mathbf{h}(\mathbf{x}^{l}_{\tau}).
\end{align}
Note that the transition condition Eqs.~\eqref{eq:6}--\eqref{eq:7} are caused by the delayed coupling between the elements. 


\begin{thebibliography}{54}%
\makeatletter
\providecommand \@ifxundefined [1]{%
 \@ifx{#1\undefined}
}%
\providecommand \@ifnum [1]{%
 \ifnum #1\expandafter \@firstoftwo
 \else \expandafter \@secondoftwo
 \fi
}%
\providecommand \@ifx [1]{%
 \ifx #1\expandafter \@firstoftwo
 \else \expandafter \@secondoftwo
 \fi
}%
\providecommand \natexlab [1]{#1}%
\providecommand \enquote  [1]{``#1''}%
\providecommand \bibnamefont  [1]{#1}%
\providecommand \bibfnamefont [1]{#1}%
\providecommand \citenamefont [1]{#1}%
\providecommand \href@noop [0]{\@secondoftwo}%
\providecommand \href [0]{\begingroup \@sanitize@url \@href}%
\providecommand \@href[1]{\@@startlink{#1}\@@href}%
\providecommand \@@href[1]{\endgroup#1\@@endlink}%
\providecommand \@sanitize@url [0]{\catcode `\\12\catcode `\$12\catcode
  `\&12\catcode `\#12\catcode `\^12\catcode `\_12\catcode `\%12\relax}%
\providecommand \@@startlink[1]{}%
\providecommand \@@endlink[0]{}%
\providecommand \url  [0]{\begingroup\@sanitize@url \@url }%
\providecommand \@url [1]{\endgroup\@href {#1}{\urlprefix }}%
\providecommand \urlprefix  [0]{URL }%
\providecommand \Eprint [0]{\href }%
\providecommand \doibase [0]{http://dx.doi.org/}%
\providecommand \selectlanguage [0]{\@gobble}%
\providecommand \bibinfo  [0]{\@secondoftwo}%
\providecommand \bibfield  [0]{\@secondoftwo}%
\providecommand \translation [1]{[#1]}%
\providecommand \BibitemOpen [0]{}%
\providecommand \bibitemStop [0]{}%
\providecommand \bibitemNoStop [0]{.\EOS\space}%
\providecommand \EOS [0]{\spacefactor3000\relax}%
\providecommand \BibitemShut  [1]{\csname bibitem#1\endcsname}%
\let\auto@bib@innerbib\@empty
\bibitem [{\citenamefont {Strogatz}(2001)}]{Strogatz2001}%
  \BibitemOpen
  \bibfield  {author} {\bibinfo {author} {\bibfnamefont {S. H.}\ \bibnamefont
  {Strogatz}},\ }\bibfield  {title} {\textit {\bibinfo {title} {{Exploring
  Complex Networks}}}}, \href {\doibase 10.1038/35065725} {\bibfield
  {journal} {\bibinfo  {journal} {Nature}\ }\textbf {\bibinfo {volume} {410}},\
  \bibinfo {pages} {268} (\bibinfo {year} {2001})}\BibitemShut {NoStop}%
\bibitem [{\citenamefont {Boccaletti}\ \textit {et~al.}(2006)\citenamefont
  {Boccaletti}, \citenamefont {Latora}, \citenamefont {Moreno}, \citenamefont
  {Chavez},\ and\ \citenamefont {Hwang}}]{Boccaletti2006}%
  \BibitemOpen
  \bibfield  {author} {\bibinfo {author} {\bibfnamefont {S.}~\bibnamefont
  {Boccaletti}}, \bibinfo {author} {\bibfnamefont {V.}~\bibnamefont {Latora}},
  \bibinfo {author} {\bibfnamefont {Y.}~\bibnamefont {Moreno}}, \bibinfo
  {author} {\bibfnamefont {M.}~\bibnamefont {Chavez}}, and\ \bibinfo {author}
  {\bibfnamefont {D.}~\bibnamefont {Hwang}},\ }\bibfield  {title} {\textit
  {\bibinfo {title} {{Complex Networks: Structure and Dynamics}}}}, \href
  {\doibase 10.1016/j.physrep.2005.10.009} {\bibfield  {journal} {\bibinfo
  {journal} {Phys. Rep.}\ }\textbf {\bibinfo {volume} {424}},\ \bibinfo {pages}
  {175} (\bibinfo {year} {2006})}\BibitemShut {NoStop}%
\bibitem [{\citenamefont {Arenas}\ \textit {et~al.}(2008)\citenamefont {Arenas},
  \citenamefont {D\'{\i}az-Guilera}, \citenamefont {Kurths}, \citenamefont
  {Moreno}, and\ \citenamefont {Zhou}}]{Arenas2008}%
  \BibitemOpen
  \bibfield  {author} {\bibinfo {author} {\bibfnamefont {A.}\ \bibnamefont
  {Arenas}}, \bibinfo {author} {\bibfnamefont {A.}\ \bibnamefont
  {D\'{\i}az-Guilera}}, \bibinfo {author} {\bibfnamefont {J.}\ \bibnamefont
  {Kurths}}, \bibinfo {author} {\bibfnamefont {Y.}\ \bibnamefont {Moreno}},
  and\ \bibinfo {author} {\bibfnamefont {C.}\ \bibnamefont {Zhou}},\
  }\bibfield  {title} {\textit {\bibinfo {title} {{Synchronization in Complex
  Networks}}}}, \href {\doibase 10.1016/j.physrep.2008.09.002} {\bibfield
  {journal} {\bibinfo  {journal} {Phys. Rep.}\ }\textbf {\bibinfo {volume}
  {469}},\ \bibinfo {pages} {93} (\bibinfo {year} {2008})}\BibitemShut
  {NoStop}%
\bibitem [{\citenamefont {Engel}\ \textit {et~al.}(2001)\citenamefont {Engel},
  \citenamefont {Fries},\ and\ \citenamefont {Singer}}]{Engel2001}%
  \BibitemOpen
  \bibfield  {author} {\bibinfo {author} {\bibfnamefont {A. K.}~\bibnamefont
  {Engel}}, \bibinfo {author} {\bibfnamefont {P.}~\bibnamefont {Fries}}, and\
  \bibinfo {author} {\bibfnamefont {W.}~\bibnamefont {Singer}},\ }\bibfield
  {title} {\textit {\bibinfo {title} {{Dynamic Predictions: Oscillations and
  Synchrony in Top-Down Processing}}}}, \href {\doibase 10.1038/35094565}
  {\bibfield  {journal} {\bibinfo  {journal} {Nat. Rev. Neurosci.}\ }\textbf
  {\bibinfo {volume} {2}},\ \bibinfo {pages} {704} (\bibinfo {year}
  {2001})}\BibitemShut {NoStop}%
\bibitem [{\citenamefont {Schnitzler}\ and\ \citenamefont
  {Gross}(2005)}]{Schnitzler2005}%
  \BibitemOpen
  \bibfield  {author} {\bibinfo {author} {\bibfnamefont {A.}\ \bibnamefont
  {Schnitzler}} and\ \bibinfo {author} {\bibfnamefont {J.}\ \bibnamefont
  {Gross}},\ }\bibfield  {title} {\textit {\bibinfo {title} {{Normal and
  Pathological Oscillatory Communication in the Brain}}}}, \href {\doibase
  10.1038/nrn1650} {\bibfield  {journal} {\bibinfo  {journal} {Nat. Rev.
  Neurosci.}\ }\textbf {\bibinfo {volume} {6}},\ \bibinfo {pages} {285}
  (\bibinfo {year} {2005})}\BibitemShut {NoStop}%
\bibitem [{\citenamefont {Uhlhaas}\ and\ \citenamefont
  {Singer}(2010)}]{Uhlhaas2010}%
  \BibitemOpen
  \bibfield  {author} {\bibinfo {author} {\bibfnamefont {P.~J.}\ \bibnamefont
  {Uhlhaas}} and\ \bibinfo {author} {\bibfnamefont {W.}\ \bibnamefont
  {Singer}},\ }\bibfield  {title} {\textit {\bibinfo {title} {{Abnormal Neural
  Oscillations and Synchrony in Schizophrenia.}}}}, \href {\doibase
  10.1038/nrn2774} {\bibfield  {journal} {\bibinfo  {journal} {Nat. Rev.
  Neurosci.}\ }\textbf {\bibinfo {volume} {11}},\ \bibinfo {pages} {100}
  (\bibinfo {year} {2010})}\BibitemShut {NoStop}%
\bibitem [{\citenamefont {Wang}(2010)}]{Wang2010}%
  \BibitemOpen
  \bibfield  {author} {\bibinfo {author} {\bibfnamefont {X.-J.}\ \bibnamefont
  {Wang}},\ }\bibfield  {title} {\textit {\bibinfo {title}
  {{Neurophysiological and Computational Principles of Cortical Rhythms in
  Cognition}}}}, \href {\doibase 10.1152/physrev.00035.2008.} {\bibfield
  {journal} {\bibinfo  {journal} {Physiol. Rev.}\ }\textbf {\bibinfo {volume}
  {90}},\ \bibinfo {pages} {1195} (\bibinfo {year} {2010})}\BibitemShut
  {NoStop}%
\bibitem [{\citenamefont {Pecora}\ and\ \citenamefont
  {Carroll}(1998)}]{Pecora1998}%
  \BibitemOpen
  \bibfield  {author} {\bibinfo {author} {\bibfnamefont {L.~M.}\
  \bibnamefont {Pecora}} and\ \bibinfo {author} {\bibfnamefont {T.~L.}\
  \bibnamefont {Carroll}},\ }\bibfield  {title} {\textit {\bibinfo {title}
  {{Master Stability Functions for Synchronized Coupled Systems}}}}, \href
  {\doibase 10.1103/PhysRevLett.80.2109} {\bibfield  {journal} {\bibinfo
  {journal} {Phys. Rev. Lett.}\ }\textbf {\bibinfo {volume} {80}},\ \bibinfo
  {pages} {2109} (\bibinfo {year} {1998})}\BibitemShut {NoStop}%
\bibitem [{\citenamefont {Barahona}\ and\ \citenamefont
  {Pecora}(2002)}]{Barahona2002}%
  \BibitemOpen
  \bibfield  {author} {\bibinfo {author} {\bibfnamefont {M.}\
  \bibnamefont {Barahona}} and\ \bibinfo {author} {\bibfnamefont {L.~M.}\
  \bibnamefont {Pecora}},\ }\bibfield  {title} {\textit {\bibinfo {title}
  {{Synchronization in Small-World Systems}}}}, \href {\doibase
  10.1103/PhysRevLett.89.054101} {\bibfield  {journal} {\bibinfo  {journal}
  {Phys. Rev. Lett.}\ }\textbf {\bibinfo {volume} {89}},\ \bibinfo {pages}
  {054101} (\bibinfo {year} {2002})}\BibitemShut {NoStop}%
\bibitem [{\citenamefont {Nishikawa}\ \textit {et~al.}(2003)\citenamefont
  {Nishikawa}, \citenamefont {Motter}, \citenamefont {Lai},\ and\ \citenamefont
  {Hoppensteadt}}]{Nishikawa2003}%
  \BibitemOpen
  \bibfield  {author} {\bibinfo {author} {\bibfnamefont {T.}\ \bibnamefont
  {Nishikawa}}, \bibinfo {author} {\bibfnamefont {A.~E.}\ \bibnamefont
  {Motter}}, \bibinfo {author} {\bibfnamefont {Y.-C.}\ \bibnamefont
  {Lai}}, and\ \bibinfo {author} {\bibfnamefont {F.~C.}\ \bibnamefont
  {Hoppensteadt}},\ }\bibfield  {title} {\textit {\bibinfo {title}
  {{Heterogeneity in Oscillator Networks: Are Smaller Worlds Easier to
  Synchronize?}}}}, \href {\doibase 10.1103/PhysRevLett.91.014101} {\bibfield
  {journal} {\bibinfo  {journal} {Phys. Rev. Lett.}\ }\textbf {\bibinfo
  {volume} {91}},\ \bibinfo {pages} {014101} (\bibinfo {year}
  {2003})}\BibitemShut {NoStop}%
\bibitem [{\citenamefont {Chavez}\ \textit {et~al.}(2005)\citenamefont {Chavez},
  \citenamefont {Hwang}, \citenamefont {Amann}, \citenamefont {Hentschel},\
  and\ \citenamefont {Boccaletti}}]{Chavez2005}%
  \BibitemOpen
  \bibfield  {author} {\bibinfo {author} {\bibfnamefont {M.}~\bibnamefont
  {Chavez}}, \bibinfo {author} {\bibfnamefont {D.-U.}\ \bibnamefont {Hwang}},
  \bibinfo {author} {\bibfnamefont {A.}~\bibnamefont {Amann}}, \bibinfo
  {author} {\bibfnamefont {H.~G.~E.}\ \bibnamefont {Hentschel}}, and\
  \bibinfo {author} {\bibfnamefont {S.}~\bibnamefont {Boccaletti}},\ }\bibfield
   {title} {\textit {\bibinfo {title} {{Synchronization is Enhanced in
  Weighted Complex Networks}}}}, \href {\doibase
  10.1103/PhysRevLett.94.218701} {\bibfield  {journal} {\bibinfo  {journal}
  {Phys. Rev. Lett.}\ }\textbf {\bibinfo {volume} {94}},\ \bibinfo {pages}
  {218701} (\bibinfo {year} {2005})}\BibitemShut {NoStop}%
\bibitem [{\citenamefont {Dhamala}\ \textit {et~al.}(2004)\citenamefont
  {Dhamala}, \citenamefont {Jirsa},\ and\ \citenamefont {Ding}}]{Dhamala2004}%
  \BibitemOpen
  \bibfield  {author} {\bibinfo {author} {\bibfnamefont {M.}\
  \bibnamefont {Dhamala}}, \bibinfo {author} {\bibfnamefont {V.}\
  \bibnamefont {Jirsa}}, and\ \bibinfo {author} {\bibfnamefont {M.}\
  \bibnamefont {Ding}},\ }\bibfield  {title} {\textit {\bibinfo {title}
  {{Enhancement of Neural Synchrony by Time Delay}}}}, \href {\doibase
  10.1103/PhysRevLett.92.074104} {\bibfield  {journal} {\bibinfo  {journal}
  {Phys. Rev. Lett.}\ }\textbf {\bibinfo {volume} {92}},\ \bibinfo {pages}
  {074104} (\bibinfo {year} {2004})}\BibitemShut {NoStop}%
\bibitem [{\citenamefont {Sorrentino}\ and\ \citenamefont {Ott}(2007)}]{SOR07}%
  \BibitemOpen
  \bibfield  {author} {\bibinfo {author} {\bibfnamefont {F.}~\bibnamefont
  {Sorrentino}} and\ \bibinfo {author} {\bibfnamefont {E.}~\bibnamefont
  {Ott}},\ }\bibfield  {title} {\textit {\bibinfo {title} {Network
  Synchronization of Groups}}}, \href@noop {} {\bibfield  {journal} {\bibinfo
  {journal} {Phys. Rev.~E}\ }\textbf {\bibinfo {volume} {76}},\ \bibinfo
  {pages} {056114} (\bibinfo {year} {2007})}\BibitemShut {NoStop}%
\bibitem [{\citenamefont {Choe}\ \textit {et~al.}(2010)\citenamefont {Choe},
  \citenamefont {Dahms}, \citenamefont {H\"{o}vel},\ and\ \citenamefont
  {Sch\"{o}ll}}]{Choe2010}%
  \BibitemOpen
  \bibfield  {author} {\bibinfo {author} {\bibfnamefont {C.-U.}\
  \bibnamefont {Choe}}, \bibinfo {author} {\bibfnamefont {T.}\ \bibnamefont
  {Dahms}}, \bibinfo {author} {\bibfnamefont {P.}\ \bibnamefont
  {H\"{o}vel}}, and\ \bibinfo {author} {\bibfnamefont {E.}\
  \bibnamefont {Sch\"{o}ll}},\ }\bibfield  {title} {\textit {\bibinfo {title}
  {{Controlling Synchrony by Delay Coupling in Networks: From In-Phase to Splay
  and Cluster States}}}}, \href {\doibase 10.1103/PhysRevE.81.025205}
  {\bibfield  {journal} {\bibinfo  {journal} {Phys. Rev. E}\ }\textbf {\bibinfo
  {volume} {81}},\ \bibinfo {pages} {025205(R)} (\bibinfo {year}
  {2010})}\BibitemShut {NoStop}%
\bibitem [{\citenamefont {Selivanov}\ \textit {et~al.}(2012)\citenamefont
  {Selivanov}, \citenamefont {Lehnert}, \citenamefont {Dahms}, \citenamefont
  {H{\"o}vel}, \citenamefont {Fradkov},\ and\ \citenamefont
  {Sch{\"o}ll}}]{SEL12}%
  \BibitemOpen
  \bibfield  {author} {\bibinfo {author} {\bibfnamefont {A.~A.}\ \bibnamefont
  {Selivanov}}, \bibinfo {author} {\bibfnamefont {J.}~\bibnamefont {Lehnert}},
  \bibinfo {author} {\bibfnamefont {T.}~\bibnamefont {Dahms}}, \bibinfo
  {author} {\bibfnamefont {P.}~\bibnamefont {H{\"o}vel}}, \bibinfo {author}
  {\bibfnamefont {A.~L.}\ \bibnamefont {Fradkov}}, and\ \bibinfo {author}
  {\bibfnamefont {E.}~\bibnamefont {Sch{\"o}ll}},\ }\bibfield  {title}
  {\textit {\bibinfo {title} {Adaptive Synchronization in Delay-Coupled
  Networks of {Stuart-Landau} Oscillators}}}, \href@noop {} {\bibfield
  {journal} {\bibinfo  {journal} {Phys. Rev.~E}\ }\textbf {\bibinfo {volume}
  {85}},\ \bibinfo {pages} {016201} (\bibinfo {year} {2012})}\BibitemShut
  {NoStop}%
\bibitem [{\citenamefont {Williams}\ \textit {et~al.}(2013)\citenamefont
  {Williams}, \citenamefont {Murphy}, \citenamefont {Roy}, \citenamefont
  {Sorrentino}, \citenamefont {Dahms},\ and\ \citenamefont
  {Sch{\"o}ll}}]{WIL13}%
  \BibitemOpen
  \bibfield  {author} {\bibinfo {author} {\bibfnamefont {C.~R.~S.}\ \bibnamefont
  {Williams}}, \bibinfo {author} {\bibfnamefont {T.~E.}\ \bibnamefont {Murphy}},
  \bibinfo {author} {\bibfnamefont {R.}~\bibnamefont {Roy}}, \bibinfo {author}
  {\bibfnamefont {F.}~\bibnamefont {Sorrentino}}, \bibinfo {author}
  {\bibfnamefont {T.}~\bibnamefont {Dahms}}, and\ \bibinfo {author}
  {\bibfnamefont {E.}~\bibnamefont {Sch{\"o}ll}},\ }\bibfield  {title} {\textit
  {\bibinfo {title} {Experimental Observations of Group Synchrony in a System of Chaotic Optoelectronic Oscillators}}}, \href@noop {} {\bibfield  {journal} {\bibinfo
  {journal} {Phys. Rev. Lett.}\ }\textbf {\bibinfo {volume} {110}},\ \bibinfo
  {pages} {064104} (\bibinfo {year} {2013})}\BibitemShut {NoStop}%
\bibitem [{\citenamefont {Dahms}\ \textit {et~al.}(2012)\citenamefont {Dahms},
  \citenamefont {Lehnert},\ and\ \citenamefont {Sch\"{o}ll}}]{DAH12}%
  \BibitemOpen
  \bibfield  {author} {\bibinfo {author} {\bibfnamefont {T.}~\bibnamefont
  {Dahms}}, \bibinfo {author} {\bibfnamefont {J.}~\bibnamefont {Lehnert}}, \
  and\ \bibinfo {author} {\bibfnamefont {E.}~\bibnamefont {Sch\"{o}ll}},\
  }\bibfield  {title} {\textit {\bibinfo {title} {{Cluster and Group
  Synchronization in Delay-Coupled Networks}}}}, \href {\doibase
  10.1103/PhysRevE.86.016202} {\bibfield  {journal} {\bibinfo  {journal} {Phys.
  Rev. E}\ }\textbf {\bibinfo {volume} {86}},\ \bibinfo {pages} {016202}
  (\bibinfo {year} {2012})}\BibitemShut {NoStop}%
\bibitem [{\citenamefont {Bernardo}\ \textit {et~al.}(2008)\citenamefont
  {Bernardo}, \citenamefont {Budd}, \citenamefont {Champneys}, \citenamefont
  {Kowalczyk}, \citenamefont {Nordmark}, \citenamefont {Tost}, \citenamefont
  {Piiroinen}, \citenamefont {Bernardo},\ and\ \citenamefont
  {Al}}]{Bernardo2008}%
  \BibitemOpen
  \bibfield  {author} {\bibinfo {author} {\bibfnamefont {M.}\ \bibnamefont
  {Di~Bernardo}}, \bibinfo {author} {\bibfnamefont {C.~J.}\ \bibnamefont
  {Budd}}, \bibinfo {author} {\bibfnamefont {A.~R.}\ \bibnamefont
  {Champneys}}, \bibinfo {author} {\bibfnamefont {P.}\ \bibnamefont
  {Kowalczyk}}, \bibinfo {author} {\bibfnamefont {A.~B.}\ \bibnamefont
  {Nordmark}}, \bibinfo {author} {\bibfnamefont {G.~O.}\ \bibnamefont
  {Tost}}, \bibinfo {author} {\bibfnamefont {P.~T.}\ \bibnamefont
  {Piiroinen}}, \bibinfo {author} {\bibfnamefont {D.~I.}\ \bibnamefont
  {Bernardo}}, and\ \bibinfo {author} {\bibfnamefont {E.~T.}\ \bibnamefont
  {Al}},\ }\bibfield  {title} {\textit {\bibinfo {title} {{Bifurcations in
  Nonsmooth Dynamical Systems}}}}, \href@noop {} {\bibfield  {journal}
  {\bibinfo  {journal} {SIAM Rev.}\ }\textbf {\bibinfo {volume} {50}},\
  \bibinfo {pages} {629} (\bibinfo {year} {2008})}\BibitemShut {NoStop}%
\bibitem [{\citenamefont {Bernardo}\ \textit {et~al.}(1998)\citenamefont
  {Bernardo}, \citenamefont {Garofalo}, \citenamefont {Glielmo},\ and\
  \citenamefont {Vasca}}]{Bernardo1998}%
  \BibitemOpen
  \bibfield  {author} {\bibinfo {author} {\bibfnamefont {M.}\ \bibnamefont
  {Di~Bernardo}}, \bibinfo {author} {\bibfnamefont {F.}\ \bibnamefont
  {Garofalo}}, \bibinfo {author} {\bibfnamefont {L.}\ \bibnamefont
  {Glielmo}}, and\ \bibinfo {author} {\bibfnamefont {F.}\ \bibnamefont
  {Vasca}},\ }\bibfield  {title} {\textit {\bibinfo {title} {{Switchings,
  Bifurcations, and Chaos in DC/DC Converters}}}}, \href@noop {} {\bibfield
   {journal} {\bibinfo  {journal} {IEEE Trans. Circuits Syst. I, Reg. Papers}\
  }\textbf {\bibinfo {volume} {45}},\ \bibinfo {pages} {133} (\bibinfo
  {year} {1998})}\BibitemShut {NoStop}%
\bibitem [{\citenamefont {Banerjee}\ and\ \citenamefont
  {Verghese}(2001)}]{Banerjee2001}%
  \BibitemOpen
  {\textit {\bibinfo {title}
  {{Nonlinear Phenomena in Power Electronics: Bifurcations, Chaos, Control, 
and Applications}}}},\ {\bibfnamefont {edited by S.}\
  \bibnamefont {Banerjee}} and\ {\bibfnamefont {G.~C.}\
  \bibnamefont {Verghese}}\ (\bibinfo  {publisher}
  {Wiley-IEEE Press, New York},\ \bibinfo {year} {2001})\BibitemShut {NoStop}%
\bibitem [{\citenamefont {Ye}\ \textit {et~al.}(1998)\citenamefont {Ye},
  \citenamefont {Michel},\ and\ \citenamefont {Hou}}]{Ye1998}%
  \BibitemOpen
  \bibfield  {author} {\bibinfo {author} {\bibfnamefont {H.}\ \bibnamefont
  {Ye}}, \bibinfo {author} {\bibfnamefont {A.~N.}\ \bibnamefont {Michel}},
  and\ \bibinfo {author} {\bibfnamefont {L.}\ \bibnamefont {Hou}},\
  }\bibfield  {title} {\textit {\bibinfo {title} {{Stability Theory for Hybrid
  Dynamical Systems}}}}, \href@noop {} {\bibfield  {journal} {\bibinfo
  {journal} {IEEE Trans. Autom. Control}\ }\textbf {\bibinfo {volume} {43}},\
  \bibinfo {pages} {461} (\bibinfo {year} {1998})}\BibitemShut {NoStop}%
\bibitem [{\citenamefont {Cassandras}\ \textit {et~al.}(2001)\citenamefont
  {Cassandras}, \citenamefont {Pepyne},\ and\ \citenamefont
  {Wardi}}]{Cassandras2001}%
  \BibitemOpen
  \bibfield  {author} {\bibinfo {author} {\bibfnamefont {C. G.}\ \bibnamefont
  {Cassandras}}, \bibinfo {author} {\bibfnamefont {D. L.}\ \bibnamefont
  {Pepyne}}, and\ \bibinfo {author} {\bibfnamefont {Y.}~\bibnamefont
  {Wardi}},\ }\bibfield  {title} {\textit {\bibinfo {title} {{Optimal Control
  of a Class of Hybrid Systems}}}}, \href {\doibase 10.1109/9.911417}
  {\bibfield  {journal} {\bibinfo  {journal} {IEEE Trans. Autom. Control}\
  }\textbf {\bibinfo {volume} {46}},\ \bibinfo {pages} {398} (\bibinfo
  {year} {2001})}\BibitemShut {NoStop}%
\bibitem [{\citenamefont {Battogtokh}\ \textit {et~al.}(2006)\citenamefont
  {Battogtokh}, \citenamefont {Aihara},\ and\ \citenamefont
  {Tyson}}]{Battogtokh2006a}%
  \BibitemOpen
  \bibfield  {author} {\bibinfo {author} {\bibfnamefont {D.}\
  \bibnamefont {Battogtokh}}, \bibinfo {author} {\bibfnamefont {K.}\
  \bibnamefont {Aihara}}, and\ \bibinfo {author} {\bibfnamefont {J.}\
  \bibnamefont {Tyson}},\ }\bibfield  {title} {\textit {\bibinfo {title}
  {{Synchronization of Eukaryotic Cells by Periodic Forcing}}}}, \href
  {\doibase 10.1103/PhysRevLett.96.148102} {\bibfield  {journal} {\bibinfo
  {journal} {Phys. Rev. Lett.}\ }\textbf {\bibinfo {volume} {96}},\ \bibinfo
  {pages} {011910} (\bibinfo {year} {2006})}\BibitemShut {NoStop}%
\bibitem [{\citenamefont {Aihara}\ and\ \citenamefont
  {Suzuki}(2010)}]{Aihara2010}%
  \BibitemOpen
  \bibfield  {author} {\bibinfo {author} {\bibfnamefont {K.}\
  \bibnamefont {Aihara}} and\ \bibinfo {author} {\bibfnamefont {H.}\
  \bibnamefont {Suzuki}},\ }\bibfield  {title} {\textit {\bibinfo {title}
  {{Theory of Hybrid Dynamical Systems and its Applications to Biological and
  Medical Systems}}}}, \href {\doibase 10.1098/rsta.2010.0237} {\bibfield
  {journal} {\bibinfo  {journal} {Phil. Trans. R. Soc. A}\ }\textbf {\bibinfo
  {volume} {368}},\ \bibinfo {pages} {4893} (\bibinfo {year}
  {2010})}\BibitemShut {NoStop}%
\bibitem [{\citenamefont {Izhikevich}(2007)}]{Izhikevich2007}%
  \BibitemOpen
  \bibfield  {author} {\bibinfo {author} {\bibfnamefont {E.~M.}\
  \bibnamefont {Izhikevich}},\ }\href@noop {} {\textit {\bibinfo {title}
  {Dynamical Systems in Neuroscience: The Geometry of Excitability and Bursting}}}\ (\bibinfo  {publisher} {MIT Press, Cambridge, MA},\ \bibinfo {year}
  {2007})\BibitemShut {NoStop}%
\bibitem [{\citenamefont {Izhikevich}\ and\ \citenamefont
  {Edelman}(2008)}]{Izhikevich2008}%
  \BibitemOpen
  \bibfield  {author} {\bibinfo {author} {\bibfnamefont {E.~M.}\
  \bibnamefont {Izhikevich}} and\ \bibinfo {author} {\bibfnamefont {G.}\
  \bibnamefont {Edelman}},\ }\bibfield  {title} {\textit {\bibinfo {title}
  {{Large-Scale Model of Mammalian Thalamocortical Systems}}}}, \href
  {\doibase 10.1073/pnas.0712231105} {\bibfield  {journal} {\bibinfo  {journal}
  {Proc. Natl. Acad. Sci. USA}\ }\textbf {\bibinfo {volume} {105}},\ \bibinfo
  {pages} {3593} (\bibinfo {year} {2008})}\BibitemShut {NoStop}%
\bibitem [{\citenamefont {Vogels}\ and\ \citenamefont
  {Abbott}(2009)}]{Vogels2009}%
  \BibitemOpen
  \bibfield  {author} {\bibinfo {author} {\bibfnamefont {T.~P.}\ \bibnamefont
  {Vogels}} and\ \bibinfo {author} {\bibfnamefont {L.~F.}\ \bibnamefont
  {Abbott}},\ }\bibfield  {title} {\textit {\bibinfo {title} {{Gating Multiple
  Signals through Detailed Balance of Excitation and Inhibition in Spiking
  Networks}}}}, \href {\doibase 10.1038/nn.2276} {\bibfield  {journal}
  {\bibinfo  {journal} {Nat. Neurosci.}\ }\textbf {\bibinfo {volume} {12}},\
  \bibinfo {pages} {483} (\bibinfo {year} {2009})}\BibitemShut {NoStop}%
\bibitem [{\citenamefont {Litwin-Kumar}\ and\ \citenamefont
  {Doiron}(2012)}]{Litwin-Kumar2012}%
  \BibitemOpen
  \bibfield  {author} {\bibinfo {author} {\bibfnamefont {A.}\ \bibnamefont
  {Litwin-Kumar}} and\ \bibinfo {author} {\bibfnamefont {B.}\ \bibnamefont
  {Doiron}},\ }\bibfield  {title} {\textit {\bibinfo {title} {{Slow Dynamics
  and High Variability in Balanced Cortical Networks with Clustered
  Connections}}}}, \href {\doibase 10.1038/nn.3220} {\bibfield  {journal}
  {\bibinfo  {journal} {Nat. Neurosci.}\ }\textbf {\bibinfo {volume} {15}},\
  \bibinfo {pages} {1498} (\bibinfo {year} {2012})}\BibitemShut {NoStop}%
\bibitem [{\citenamefont {Destexhe}(2009)}]{Destexhe2009}%
  \BibitemOpen
  \bibfield  {author} {\bibinfo {author} {\bibfnamefont {A.}\ \bibnamefont
  {Destexhe}},\ }\bibfield  {title} {\textit {\bibinfo {title}
  {{Self-Sustained Asynchronous Irregular States and Up-Down States in
  Thalamic, Cortical and Thalamocortical Networks of Nonlinear
  Integrate-and-Fire Neurons}}}}, \href {\doibase 10.1007/s10827-009-0164-4}
  {\bibfield  {journal} {\bibinfo  {journal} {J. Comput. Neurosci.}\ }\textbf
  {\bibinfo {volume} {27}},\ \bibinfo {pages} {493} (\bibinfo {year}
  {2009})}\BibitemShut {NoStop}%
\bibitem [{\citenamefont {Brunel}(2000)}]{Brunel2000}%
  \BibitemOpen
  \bibfield  {author} {\bibinfo {author} {\bibfnamefont {N.}~\bibnamefont
  {Brunel}},\ }\bibfield  {title} {\textit {\bibinfo {title} {{Dynamics of
  Sparsely Connected Networks of Excitatory and Inhibitory Spiking Neurons}}}}, \href@noop {} {\bibfield  {journal} {\bibinfo  {journal} {J. Comput.
  Neurosci.}\ }\textbf {\bibinfo {volume} {8}},\ \bibinfo {pages} {183}
  (\bibinfo {year} {2000})}\BibitemShut {NoStop}%
\bibitem [{\citenamefont {Gigante}\ \textit {et~al.}(2007)\citenamefont
  {Gigante}, \citenamefont {Mattia},\ and\ \citenamefont
  {Giudice}}]{Gigante2007c}%
  \BibitemOpen
  \bibfield  {author} {\bibinfo {author} {\bibfnamefont {G.}\ \bibnamefont
  {Gigante}}, \bibinfo {author} {\bibfnamefont {M.}\ \bibnamefont
  {Mattia}}, and\ \bibinfo {author} {\bibfnamefont {P.}\ \bibnamefont
  {Giudice}},\ }\bibfield  {title} {\textit {\bibinfo {title} {{Diverse
  Population-Bursting Modes of Adapting Spiking Neurons}}}}, \href {\doibase
  10.1103/PhysRevLett.98.148101} {\bibfield  {journal} {\bibinfo  {journal}
  {Phys. Rev. Lett.}\ }\textbf {\bibinfo {volume} {98}},\ \bibinfo {pages}
  {148101} (\bibinfo {year} {2007})}\BibitemShut {NoStop}%
\bibitem [{\citenamefont {Augustin}\ \textit {et~al.}(2013)\citenamefont
  {Augustin}, \citenamefont {Ladenbauer},\ and\ \citenamefont
  {Obermayer}}]{Augustin2013}%
  \BibitemOpen
  \bibfield  {author} {\bibinfo {author} {\bibfnamefont {M.}\ \bibnamefont
  {Augustin}}, \bibinfo {author} {\bibfnamefont {J.}\ \bibnamefont
  {Ladenbauer}}, and\ \bibinfo {author} {\bibfnamefont {K.}\ \bibnamefont
  {Obermayer}},\ }\bibfield  {title} {\textit {\bibinfo {title} {{How
  Adaptation Shapes Spike Rate Oscillations in Recurrent Neuronal Networks}}}}, \href {\doibase 10.3389/fncom.2013.00009} {\bibfield  {journal} {\bibinfo
  {journal} {Front. Comput. Neurosci.}\ }\textbf {\bibinfo {volume} {7}},\
  \bibinfo {pages} {1} (\bibinfo {year} {2013})}\BibitemShut {NoStop}%
\bibitem [{\citenamefont {Indiveri}\ \textit {et~al.}(2006)\citenamefont
  {Indiveri}, \citenamefont {Chicca},\ and\ \citenamefont
  {Douglas}}]{Indiveri2006}%
  \BibitemOpen
  \bibfield  {author} {\bibinfo {author} {\bibfnamefont {G.}\ \bibnamefont
  {Indiveri}}, \bibinfo {author} {\bibfnamefont {E.}\ \bibnamefont
  {Chicca}}, and\ \bibinfo {author} {\bibfnamefont {R.}\ \bibnamefont
  {Douglas}},\ }\bibfield  {title} {\textit {\bibinfo {title} {{A VLSI Array
  of Low-Power Spiking Neurons and Bistable Synapses with Spike-Timing
  Dependent Plasticity}}}}, \href {\doibase 10.1109/TNN.2005.860850}
  {\bibfield  {journal} {\bibinfo  {journal} {IEEE Trans. Neural Netw.}\
  }\textbf {\bibinfo {volume} {17}},\ \bibinfo {pages} {211} (\bibinfo
  {year} {2006})}\BibitemShut {NoStop}%
\bibitem [{\citenamefont {Jo}\ \textit {et~al.}(2010)\citenamefont {Jo},
  \citenamefont {Chang}, \citenamefont {Ebong}, \citenamefont {Bhadviya},
  \citenamefont {Mazumder},\ and\ \citenamefont {Lu}}]{Jo2010}%
  \BibitemOpen
  \bibfield  {author} {\bibinfo {author} {\bibfnamefont {S.~H.}\
  \bibnamefont {Jo}}, \bibinfo {author} {\bibfnamefont {T.}\ \bibnamefont
  {Chang}}, \bibinfo {author} {\bibfnamefont {I.}\ \bibnamefont
  {Ebong}}, \bibinfo {author} {\bibfnamefont {B.~B.}\ \bibnamefont
  {Bhadviya}}, \bibinfo {author} {\bibfnamefont {P.}\ \bibnamefont
  {Mazumder}}, and\ \bibinfo {author} {\bibfnamefont {W.}\ \bibnamefont
  {Lu}},\ }\bibfield  {title} {\textit {\bibinfo {title} {{Nanoscale Memristor
  Device as Synapse in Neuromorphic Systems}}}}, \href {\doibase
  10.1021/nl904092h} {\bibfield  {journal} {\bibinfo  {journal} {Nano Lett.}\
  }\textbf {\bibinfo {volume} {10}},\ \bibinfo {pages} {1297} (\bibinfo
  {year} {2010})}\BibitemShut {NoStop}%
\bibitem [{\citenamefont {Brette}\ and\ \citenamefont
  {Gerstner}(2005)}]{Brette2005}%
  \BibitemOpen
  \bibfield  {author} {\bibinfo {author} {\bibfnamefont {R.}\ \bibnamefont
  {Brette}} and\ \bibinfo {author} {\bibfnamefont {W.}\ \bibnamefont
  {Gerstner}},\ }\bibfield  {title} {\textit {\bibinfo {title} {{Adaptive
  Exponential Integrate-and-Fire Model as an Effective Description of Neuronal
  Activity}}}}, \href {\doibase 10.1152/jn.00686.2005} {\bibfield  {journal}
  {\bibinfo  {journal} {J. Neurophysiol.}\ }\textbf {\bibinfo {volume} {94}},\
  \bibinfo {pages} {3637} (\bibinfo {year} {2005})}\BibitemShut {NoStop}%
\bibitem [{\citenamefont {Touboul}\ and\ \citenamefont
  {Brette}(2008)}]{Touboul2008}%
  \BibitemOpen
  \bibfield  {author} {\bibinfo {author} {\bibfnamefont {J.}\
  \bibnamefont {Touboul}} and\ \bibinfo {author} {\bibfnamefont {R.}\
  \bibnamefont {Brette}},\ }\bibfield  {title} {\textit {\bibinfo {title}
  {{Dynamics and Bifurcations of the Adaptive Exponential Integrate-and-Fire
  Model}}}}, \href {\doibase 10.1007/s00422-008-0267-4} {\bibfield  {journal}
  {\bibinfo  {journal} {Biol. Cybern.}\ }\textbf {\bibinfo {volume} {99}},\
  \bibinfo {pages} {319} (\bibinfo {year} {2008})}\BibitemShut {NoStop}%
\bibitem [{\citenamefont {Naud}\ \textit {et~al.}(2008)\citenamefont {Naud},
  \citenamefont {Marcille}, \citenamefont {Clopath},\ and\ \citenamefont
  {Gerstner}}]{Naud2008}%
  \BibitemOpen
  \bibfield  {author} {\bibinfo {author} {\bibfnamefont {R.}\ \bibnamefont
  {Naud}}, \bibinfo {author} {\bibfnamefont {N.}\ \bibnamefont
  {Marcille}}, \bibinfo {author} {\bibfnamefont {C.}\ \bibnamefont
  {Clopath}}, and\ \bibinfo {author} {\bibfnamefont {W.}\ \bibnamefont
  {Gerstner}},\ }\bibfield  {title} {\textit {\bibinfo {title} {{Firing
  Patterns in the Adaptive Exponential Integrate-and-Fire Model}}}}, \href
  {\doibase 10.1007/s00422-008-0264-7} {\bibfield  {journal} {\bibinfo
  {journal} {Biol. Cybern.}\ }\textbf {\bibinfo {volume} {99}},\ \bibinfo
  {pages} {335} (\bibinfo {year} {2008})}\BibitemShut {NoStop}%
\bibitem [{\citenamefont {Jolivet}\ \textit {et~al.}(2008)\citenamefont
  {Jolivet}, \citenamefont {Sch\"{u}rmann}, \citenamefont {Berger},
  \citenamefont {Naud}, \citenamefont {Gerstner},\ and\ \citenamefont
  {Roth}}]{Jolivet2008}%
  \BibitemOpen
  \bibfield  {author} {\bibinfo {author} {\bibfnamefont {R.}\ \bibnamefont
  {Jolivet}}, \bibinfo {author} {\bibfnamefont {F.}\ \bibnamefont
  {Sch\"{u}rmann}}, \bibinfo {author} {\bibfnamefont {T.~K.}\ \bibnamefont
  {Berger}}, \bibinfo {author} {\bibfnamefont {R.}\ \bibnamefont {Naud}},
  \bibinfo {author} {\bibfnamefont {W.}\ \bibnamefont {Gerstner}}, and\
  \bibinfo {author} {\bibfnamefont {A.}\ \bibnamefont {Roth}},\ }\bibfield
  {title} {\textit {\bibinfo {title} {{The Quantitative Single-Neuron Modeling
  Competition}}}}, \href {\doibase 10.1007/s00422-008-0261-x} {\bibfield
  {journal} {\bibinfo  {journal} {Biol. Cybern.}\ }\textbf {\bibinfo {volume}
  {99}},\ \bibinfo {pages} {417} (\bibinfo {year} {2008})}\BibitemShut
  {NoStop}%
\bibitem [{Note1()}]{Note1}%
  \BibitemOpen
  \bibinfo {note} {The fast sodium current of the aEIF model includes the
  effect of the sodium current, which is responsible for the generation of
  spikes.}\BibitemShut {Stop}%
\bibitem [{\citenamefont {Ermentrout}\ \textit {et~al.}(2001)\citenamefont
  {Ermentrout}, \citenamefont {Pascal},\ and\ \citenamefont
  {Gutkin}}]{Ermentrout2001}%
  \BibitemOpen
  \bibfield  {author} {\bibinfo {author} {\bibfnamefont {G.~B.}\ \bibnamefont
  {Ermentrout}}, \bibinfo {author} {\bibfnamefont {M.}\ \bibnamefont
  {Pascal}}, and\ \bibinfo {author} {\bibfnamefont {B.~S.}\ \bibnamefont
  {Gutkin}},\ }\bibfield  {title} {\textit {\bibinfo {title} {{The Effects of
  Spike Frequency Adaptation and Negative Feedback on the Synchronization of
  Neural Oscillators}}}}, \href@noop {} {\bibfield  {journal} {\bibinfo
  {journal} {Neural Comput.}\ }\textbf {\bibinfo {volume} {13}},\ \bibinfo
  {pages} {1285--1310} (\bibinfo {year} {2001})}\BibitemShut {NoStop}%
\bibitem [{\citenamefont {Ladenbauer}\ \textit {et~al.}(2012)\citenamefont
  {Ladenbauer}, \citenamefont {Augustin}, \citenamefont {Shiau},\ and\
  \citenamefont {Obermayer}}]{Ladenbauer2012}%
  \BibitemOpen
  \bibfield  {author} {\bibinfo {author} {\bibfnamefont {J.}\ \bibnamefont
  {Ladenbauer}}, \bibinfo {author} {\bibfnamefont {M.}\ \bibnamefont
  {Augustin}}, \bibinfo {author} {\bibfnamefont {L.}\ \bibnamefont
  {Shiau}}, and\ \bibinfo {author} {\bibfnamefont {K.}\ \bibnamefont
  {Obermayer}},\ }\bibfield  {title} {\textit {\bibinfo {title} {{Impact of
  Adaptation Currents on Synchronization of Coupled Exponential
  Integrate-and-Fire Neurons}}}}, \href {\doibase
  10.1371/journal.pcbi.1002478} {\bibfield  {journal} {\bibinfo  {journal}
  {PLoS Comput. Biol.}\ }\textbf {\bibinfo {volume} {8}},\ \bibinfo {pages}
  {e1002478} (\bibinfo {year} {2012})}\BibitemShut {NoStop}%
\bibitem [{\citenamefont {Kilpatrick}\ and\ \citenamefont
  {Ermentrout}(2011)}]{Kilpatrick2011}%
  \BibitemOpen
  \bibfield  {author} {\bibinfo {author} {\bibfnamefont {Z.~P.}\
  \bibnamefont {Kilpatrick}} and\ \bibinfo {author} {\bibfnamefont {G.~B.}\
  \bibnamefont {Ermentrout}},\ }\bibfield  {title} {\textit {\bibinfo {title}
  {{Sparse Gamma Rhythms Arising through Clustering in Adapting Neuronal
  Networks}}}}, \href {\doibase 10.1371/journal.pcbi.1002281} {\bibfield
  {journal} {\bibinfo  {journal} {PLoS Comput. Biol.}\ }\textbf {\bibinfo
  {volume} {7}},\ \bibinfo {pages} {e1002281} (\bibinfo {year}
  {2011})}\BibitemShut {NoStop}%
\bibitem [{Note2()}]{Note2}%
  \BibitemOpen
  \bibinfo {note} {Note that $\protect \mathbf {h}$ does not depend on
  $\protect \mathbf {x}_i$ in Eq.~\protect \textup {\hbox {\mathsurround \z@
  \protect \normalfont (\ignorespaces \ref {eq_network_comp1}\unskip
  \@@italiccorr )}}. The following derivation, however, can be extended to
  coupling functions $\protect \mathbf {h}(\protect \mathbf {x}_i,\protect
  \mathbf {x}_{j,\tau })$ in a straightforward way}\BibitemShut {NoStop}%
\bibitem [{\citenamefont {Farmer}(1982)}]{Farmer1982}%
  \BibitemOpen
  \bibfield  {author} {\bibinfo {author} {\bibfnamefont {J.~D.}\ \bibnamefont
  {Farmer}},\ }\bibfield  {title} {\textit {\bibinfo {title} {{Chaotic
  Attractors of an Infinite-Dimensional Dynamical System}}}}, \href@noop {}
  {\bibfield  {journal} {\bibinfo  {journal} {Physica D}\ }\textbf {\bibinfo
  {volume} {4}},\ \bibinfo {pages} {366} (\bibinfo {year}
  {1982})}\BibitemShut {NoStop}%
\bibitem [{Note3()}]{Note3}%
  \BibitemOpen
  \bibinfo {note} {We observed this behavior for a wide range of model
  parameters}\BibitemShut {NoStop}%
\bibitem [{\citenamefont {McCormick}(1992)}]{McCormick1992}%
  \BibitemOpen
  \bibfield  {author} {\bibinfo {author} {\bibfnamefont {D.~A.}\ \bibnamefont
  {McCormick}},\ }\bibfield  {title} {\textit {\bibinfo {title}
  {{Neurotransmitter Actions in the Thalamus and Cerebral Cortex and their Role
  in Thalamocortical Activity}}}}, \href@noop {} {\bibfield  {journal}
  {\bibinfo  {journal} {Prog. Neurobiol.}\ }\textbf {\bibinfo {volume} {39}},\
  \bibinfo {pages} {337} (\bibinfo {year} {1992})}\BibitemShut {NoStop}%
\bibitem [{\citenamefont {Abbott}\ and\ \citenamefont
  {Nelson}(2000)}]{Abbott2000}%
  \BibitemOpen
  \bibfield  {author} {\bibinfo {author} {\bibfnamefont {L.~F.}\ \bibnamefont
  {Abbott}} and\ \bibinfo {author} {\bibfnamefont {S.~B.}\ \bibnamefont
  {Nelson}},\ }\bibfield  {title} {\textit {\bibinfo {title} {{Synaptic
  Plasticity: Taming the Beast}}}}, \href {\doibase 10.1038/81453} {\bibfield
  {journal} {\bibinfo  {journal} {Nat. Neurosci.}\ }\textbf {\bibinfo {volume}
  {3}},\ \bibinfo {pages} {1178} (\bibinfo {year} {2000})}\BibitemShut
  {NoStop}%
\bibitem [{\citenamefont {Destexhe}\ and\ \citenamefont
  {Marder}(2004)}]{Destexhe2004}%
  \BibitemOpen
  \bibfield  {author} {\bibinfo {author} {\bibfnamefont {A.}\ \bibnamefont
  {Destexhe}} and\ \bibinfo {author} {\bibfnamefont {E.}\ \bibnamefont
  {Marder}},\ }\bibfield  {title} {\textit {\bibinfo {title} {{Plasticity in
  Single Neuron and Circuit Computations}}}}, \href@noop {} {\bibfield
  {journal} {\bibinfo  {journal} {Nature}\ }\textbf {\bibinfo {volume} {431}},\
  \bibinfo {pages} {789} (\bibinfo {year} {2004})}\BibitemShut {NoStop}%
\bibitem [{\citenamefont {Gray}\ and\ \citenamefont
  {Robinson}(2009)}]{Gray2009b}%
  \BibitemOpen
  \bibfield  {author} {\bibinfo {author} {\bibfnamefont {R.~T.}\
  \bibnamefont {Gray}} and\ \bibinfo {author} {\bibfnamefont {P.~A.}\
  \bibnamefont {Robinson}},\ }\bibfield  {title} {\textit {\bibinfo {title}
  {{Stability and Structural Constraints of Random Brain Networks with
  Excitatory and Inhibitory Neural Populations}}}}, \href {\doibase
  10.1007/s10827-008-0128-0} {\bibfield  {journal} {\bibinfo  {journal} {J.
  Comput. Neurosci.}\ }\textbf {\bibinfo {volume} {27}},\ \bibinfo {pages}
  {81} (\bibinfo {year} {2009})}\BibitemShut {NoStop}%
\bibitem [{\citenamefont {Rajan}\ and\ \citenamefont
  {Abbott}(2006)}]{Rajan2006}%
  \BibitemOpen
  \bibfield  {author} {\bibinfo {author} {\bibfnamefont {K.}\ \bibnamefont
  {Rajan}} and\ \bibinfo {author} {\bibfnamefont {L.}~\bibnamefont {Abbott}},\
  }\bibfield  {title} {\textit {\bibinfo {title} {{Eigenvalue Spectra of
  Random Matrices for Neural Networks}}}}, \href {\doibase
  10.1103/PhysRevLett.97.188104} {\bibfield  {journal} {\bibinfo  {journal}
  {Phys. Rev. Lett.}\ }\textbf {\bibinfo {volume} {97}},\ \bibinfo {pages}
  {188104} (\bibinfo {year} {2006})}\BibitemShut {NoStop}%
\bibitem [{Note4()}]{Note4}%
  \BibitemOpen
  \bibinfo {note} {This follows from Gershgorin's circle theorem, because the
  coupling matrices $\protect \mathbf {C}^{kl}$ ($k,l \in \protect \{\protect
  \mathcal {E},\protect \mathcal {I}\protect \}$) all have non-negative entries
  with unity row sum. Excitatory and inhibitory interactions are generated via
  the overall coupling parameters $\lambda ^{k\protect \mathcal {E}} > 0$ and
  $\lambda ^{k\protect \mathcal {I}} < 0$}\BibitemShut {NoStop}%
\bibitem [{\citenamefont {C\^{a}teau}\ \emph {et~al.}(2008)\citenamefont
  {C\^{a}teau}, \citenamefont {Kitano},\ and\ \citenamefont
  {Fukai}}]{Cateau2008}%
  \BibitemOpen
  \bibfield  {author} {\bibinfo {author} {\bibfnamefont {H.}\
  \bibnamefont {C\^{a}teau}}, \bibinfo {author} {\bibfnamefont {K.}\
  \bibnamefont {Kitano}}, and \bibinfo {author} {\bibfnamefont {T.}\
  \bibnamefont {Fukai}}, }\bibfield  {title} {\textit {\bibinfo {title}
  {{Interplay between a Phase Response Curve and Spike-Timing-Dependent
  Plasticity Leading to Wireless Clustering}}}}, \href {\doibase
  10.1103/PhysRevE.77.051909} {\bibfield  {journal} {\bibinfo  {journal} {Phys.
  Rev. E}\ }\textbf {\bibinfo {volume} {77}},\ \bibinfo {pages} {051909}
  (\bibinfo {year} {2008})}\BibitemShut {NoStop}%
\bibitem [{\citenamefont {Lubenov}\ and\ \citenamefont
  {Siapas}(2008)}]{Lubenov2008}%
  \BibitemOpen
  \bibfield  {author} {\bibinfo {author} {\bibfnamefont {E.~V.}\
  \bibnamefont {Lubenov}} and \bibinfo {author} {\bibfnamefont
  {A.~G.}\ \bibnamefont {Siapas}}, }\bibfield  {title} {\textit
  {\bibinfo {title} {{Decoupling through Synchrony in Neuronal Circuits with
  Propagation Delays}}}}, \href {\doibase 10.1016/j.neuron.2008.01.036}
  {\bibfield  {journal} {\bibinfo  {journal} {Neuron}\ }\textbf {\bibinfo
  {volume} {58}},\ \bibinfo {pages} {118} (\bibinfo {year}
  {2008})}\BibitemShut {NoStop}%
\bibitem [{\citenamefont {Hallett}(2000)}]{Hallett2000}%
  \BibitemOpen
  \bibfield  {author} {\bibinfo {author} {\bibfnamefont {M.}~\bibnamefont
  {Hallett}},\ }\bibfield  {title} {\textit {\bibinfo {title} {{Transcranial
  Magnetic Stimulation and the Human Brain}}}}, \href {\doibase
  10.1038/35018000} {\bibfield  {journal} {\bibinfo  {journal} {Nature}\
  }\textbf {\bibinfo {volume} {406}},\ \bibinfo {pages} {147} (\bibinfo
  {year} {2000})}\BibitemShut {NoStop}%
\bibitem [{\citenamefont {Kringelbach}\ \emph {et~al.}(2007)\citenamefont
  {Kringelbach}, \citenamefont {Jenkinson}, \citenamefont {Owen},\ and\
  \citenamefont {Aziz}}]{Kringelbach2007}%
  \BibitemOpen
  \bibfield  {author} {\bibinfo {author} {\bibfnamefont {M.~L.}
  \bibnamefont {Kringelbach}}, \bibinfo {author} {\bibfnamefont {N.}
  \bibnamefont {Jenkinson}}, \bibinfo {author} {\bibfnamefont {S.}
  \bibnamefont {Owen}}, and \bibinfo {author} {\bibfnamefont {T.}
  \bibnamefont {Aziz}}, }\bibfield  {title} {\textit {\bibinfo {title}
  {{Translational Principles of Deep Brain Stimulation}}}}, \href {\doibase
  10.1038/nrn2196} {\bibfield  {journal} {\bibinfo  {journal} {Nat. Rev.
  Neurosci.}\ }\textbf {\bibinfo {volume} {8}},\ \bibinfo {pages} {623}
  (\bibinfo {year} {2007})}\BibitemShut {NoStop}%
\bibitem [{\citenamefont {Esser}\ \emph {et~al.}(2005)\citenamefont {Esser},
  \citenamefont {Hill},\ and\ \citenamefont {Tononi}}]{Esser2005}%
  \BibitemOpen
  \bibfield  {author} {\bibinfo {author} {\bibfnamefont {S.~K.} \bibnamefont
  {Esser}}, \bibinfo {author} {\bibfnamefont {S.~L.}\ \bibnamefont {Hill}},
  and \bibinfo {author} {\bibfnamefont {G.} \bibnamefont {Tononi}},
  }\bibfield  {title} {\textit {\bibinfo {title} {{Modeling the Effects of
  Transcranial Magnetic Stimulation on Cortical Circuits}}}}, \href {\doibase
  10.1152/jn.01230.2004} {\bibfield  {journal} {\bibinfo  {journal} {J.
  Neurophysiol.}\ }\textbf {\bibinfo {volume} {94}},\ \bibinfo {pages}
  {622} (\bibinfo {year} {2005})}\BibitemShut {NoStop}%
\bibitem [{\citenamefont {Tass}(2003)}]{Tass2003}%
  \BibitemOpen
  \bibfield  {author} {\bibinfo {author} {\bibfnamefont {P.} \bibnamefont
  {Tass}}, }\bibfield  {title} {\textit {\bibinfo {title} {{A Model of
  Desynchronizing Deep Brain Stimulation with a Demand-Controlled Coordinated
  Reset of Neural Subpopulations}}}}, \href {\doibase
  10.1007/s00422-003-0425-7} {\bibfield  {journal} {\bibinfo  {journal} {Biol.
  Cybern.} }\textbf {\bibinfo {volume} {89}},\ \bibinfo {pages} {81}
  (\bibinfo {year} {2003})}\BibitemShut {NoStop}%
\bibitem [{\citenamefont {Perkins}\ \emph {et~al.}(2010)\citenamefont
  {Perkins}, \citenamefont {Wilds},\ and\ \citenamefont {Glass}}]{Perkins2010}%
  \BibitemOpen
  \bibfield  {author} {\bibinfo {author} {\bibfnamefont {T.~J.}
  \bibnamefont {Perkins}}, \bibinfo {author} {\bibfnamefont {R.} \bibnamefont
  {Wilds}}, and \bibinfo {author} {\bibfnamefont {L.} \bibnamefont
  {Glass}}, }\bibfield  {title} {\textit {\bibinfo {title} {{Robust Dynamics
  in Minimal Hybrid Models of Genetic Networks}}}}, \href {\doibase
  10.1098/rsta.2010.0139} {\bibfield  {journal} {\bibinfo  {journal} {Phil.
  Trans. R. Soc. A}\ }\textbf {\bibinfo {volume} {368}},\ \bibinfo {pages}
  {4961} (\bibinfo {year} {2010})}\BibitemShut {NoStop}%
\bibitem [{\citenamefont {Zhang}\ \emph {et~al.}(2011)\citenamefont {Zhang},
  \citenamefont {Tang}, \citenamefont {Fang},\ and\ \citenamefont
  {Zhu}}]{Zhang2011}%
  \BibitemOpen
  \bibfield  {author} {\bibinfo {author} {\bibfnamefont {W.} \bibnamefont
  {Zhang}}, \bibinfo {author} {\bibfnamefont {Y.} \bibnamefont {Tang}},
  \bibinfo {author} {\bibfnamefont {J.} \bibnamefont {Fang}}, and
  \bibinfo {author} {\bibfnamefont {W.}~\bibnamefont {Zhu}}, }\bibfield
  {title} {\textit {\bibinfo {title} {{Exponential Cluster Synchronization of
  Impulsive Delayed Genetic Oscillators with External Disturbances}}}}, \href
  {\doibase 10.1063/1.3671609} {\bibfield  {journal} {\bibinfo  {journal}
  {Chaos} }\textbf {\bibinfo {volume} {21}},\ \bibinfo {pages} {043137}
  (\bibinfo {year} {2011})}\BibitemShut {NoStop}%
\bibitem [{\citenamefont {{De Lellis}}\ \emph {et~al.}(2008)\citenamefont {{De
  Lellis}}, \citenamefont {di~Bernardo},\ and\ \citenamefont
  {Garofalo}}]{DeLellis2008}%
  \BibitemOpen
  \bibfield  {author} {\bibinfo {author} {\bibfnamefont {P.} \bibnamefont
  {{De Lellis}}}, \bibinfo {author} {\bibfnamefont {M.} \bibnamefont
  {Di~Bernardo}}, and \bibinfo {author} {\bibfnamefont {F.}
  \bibnamefont {Garofalo}}, }\bibfield  {title} {\textit {\bibinfo {title}
  {{Synchronization of Complex Networks through Local Adaptive Coupling}}}},
  \href {\doibase 10.1063/1.2944236} {\bibfield  {journal} {\bibinfo
  {journal} {Chaos}\ }\textbf {\bibinfo {volume} {18}},\ \bibinfo {pages}
  {037110} (\bibinfo {year} {2008})}\BibitemShut {NoStop}%
\bibitem [{\citenamefont {Amann}\ \textit {et~al.}(2003)\citenamefont {Amann},
  \citenamefont {Peters}, \citenamefont {Parlitz}, \citenamefont {Wacker},\
  and\ \citenamefont {Sch\"{o}ll}}]{AMA03}%
  \BibitemOpen
  \bibfield  {author} {\bibinfo {author} {\bibfnamefont {A.}~\bibnamefont
  {Amann}}, \bibinfo {author} {\bibfnamefont {K.}~\bibnamefont {Peters}},
  \bibinfo {author} {\bibfnamefont {U.}~\bibnamefont {Parlitz}}, \bibinfo
  {author} {\bibfnamefont {A.}~\bibnamefont {Wacker}}, and\ \bibinfo {author}
  {\bibfnamefont {E.}~\bibnamefont {Sch\"{o}ll}},\ }\bibfield  {title}
  {\textit {\bibinfo {title} {{Hybrid Model for Chaotic Front Dynamics: From
  Semiconductors to Water Tanks}}}}, \href@noop {} {\bibfield  {journal}
  {\bibinfo  {journal} {Phys.~Rev.~Lett.}\ }\textbf {\bibinfo {volume} {91}},\
  \bibinfo {pages} {066601} (\bibinfo {year} {2003})}\BibitemShut
  {NoStop}%
\bibitem [{\citenamefont {Soriano}\ \textit {et~al.}(2013)\citenamefont
  {Soriano}, \citenamefont {García-Ojalvo}, \citenamefont {Mirasso},\ and\
  \citenamefont {Fischer}}]{SOR13}%
  \BibitemOpen
  \bibfield  {author} {\bibinfo {author} {\bibfnamefont {M.~C.}\ \bibnamefont
  {Soriano}}, \bibinfo {author} {\bibfnamefont {J.}~\bibnamefont
  {García-Ojalvo}}, \bibinfo {author} {\bibfnamefont {C.~R.}\ \bibnamefont
  {Mirasso}}, and\ \bibinfo {author} {\bibfnamefont {I.}~\bibnamefont
  {Fischer}},\ }\bibfield  {title} {\textit {\bibinfo {title} {{Complex
  Photonics: Dynamics and Applications of Delay-Coupled Semiconductors
  Lasers}}}}, \href@noop {} {\bibfield  {journal} {\bibinfo  {journal}
  {Rev.~Mod.~Phys.}\ }\textbf {\bibinfo {volume} {85}},\ \bibinfo {pages}
  {421} (\bibinfo {year} {2013})}\BibitemShut {NoStop}%
\bibitem [{\citenamefont {Deco}\ \textit {et~al.}(2011)\citenamefont {Deco},
  \citenamefont {Jirsa},\ and\ \citenamefont {McIntosh}}]{Deco2011}%
  \BibitemOpen
  \bibfield  {author} {\bibinfo {author} {\bibfnamefont {G.}\ \bibnamefont
  {Deco}}, \bibinfo {author} {\bibfnamefont {V.}\ \bibnamefont {Jirsa}},
  and\ \bibinfo {author} {\bibfnamefont {A.~R.}\ \bibnamefont
  {McIntosh}},\ }\bibfield  {title} {\textit {\bibinfo {title} {{Emerging
  Concepts for the Dynamical Organization of Resting-State Activity in the
  Brain}}}}, \href {\doibase 10.1038/nrn2961} {\bibfield  {journal} {\bibinfo
  {journal} {Nat. Rev. Neurosci.}\ }\textbf {\bibinfo {volume} {12}},\ \bibinfo
  {pages} {43} (\bibinfo {year} {2011})}\BibitemShut {NoStop}%
\bibitem [{\citenamefont {Mueller}(1995)}]{Mueller1995}%
  \BibitemOpen
  \bibfield  {author} {\bibinfo {author} {\bibfnamefont {P.~C.}\ \bibnamefont
  {Mueller}}, }\bibfield  {title} {\textit {\bibinfo {title} {{Calculation of Lyapunov Exponents for Dynamic Systems with Discontinuities}}}},\ \href@noop {} {\bibfield  {journal} {\bibinfo  {journal} {Chaos
  Soliton. Fract.}\ }\textbf {\bibinfo {volume} {5}},\ \bibinfo {pages} {1671}
  (\bibinfo {year} {1995})}\BibitemShut {NoStop}%
\end{thebibliography}
%

\end{document}